\documentclass{article}

\usepackage{arxiv}

\usepackage[utf8]{inputenc} 
\usepackage[T1]{fontenc}    
\usepackage{hyperref}       
\usepackage{url}            
\usepackage{booktabs}       
\usepackage{nicefrac}       
\usepackage{microtype}      
\usepackage{lipsum}		
\usepackage{natbib}
\usepackage{doi}

\usepackage{cite}
\usepackage{amsmath,amssymb,amsfonts}
\usepackage{bm} 
\usepackage{graphicx}
\usepackage{caption}
\usepackage{multirow}

\usepackage{tabularx} 

\usepackage{subfigure}

\usepackage{cleveref}

\usepackage{makecell}

\usepackage{algorithm,algpseudocode}
\algnewcommand{\Inputs}[1]{%
  \State \textbf{Inputs:}
  \Statex \hspace*{\algorithmicindent}\parbox[t]{.8\linewidth}{\raggedright #1}
}
\algnewcommand{\Initialize}[1]{%
  \State \textbf{Initialize:}
  \Statex \hspace*{\algorithmicindent}\parbox[t]{.8\linewidth}{\raggedright #1}
}

\usepackage{xcolor}

\newtheorem{theorem}{Theorem}
\newtheorem{lemma}{Lemma}
\newtheorem{proof}{Proof}

\title{A UCB-based Tree Search Approach to Joint Verification-Correction Strategy for Large Scale Systems}

\author{ {Peng~Xu} \\
	Grado Department of Industrial and Systems Engineering\\
	Virginia Tech\\
	Blacskburg, VA 24060 \\
	\texttt{xupeng@vt.edu} \\
	\And
	{Xinwei~Deng} \\
	Department of Statistics\\
	Virginia Tech\\
	Blacskburg, VA 24060 \\
	\texttt{xdeng@vt.edu} \\
	\And
	{Alejandro~Salado} \\
	Department of Systems and Industrial Engineering\\
	The University of Arizona\\
	Tucson, AZ 85721 \\
	\texttt{alejandrosalado@arizona.edu} \\
}

\renewcommand{\shorttitle}{ }

\hypersetup{
pdftitle={A UCB-based Tree Search Approach to Joint Verification-Correction Strategy for Large Scale Systems},
pdfsubject={EESS},
pdfauthor={Peng~Xu, Xinwei~Deng, and~Alejandro~Salado},
pdfkeywords={verification planning, Bayesian network, multi-armed bandit problem, correction activity, tree search, random forest},
}

\begin{document}
\maketitle

\begin{abstract}
	Verification planning is a sequential decision-making problem that specifies a set of verification activities (VA) and correction activities (CA) at different phases of system development. While VAs are used to identify errors and defects, CAs also play important roles in system verification as they correct the identified errors and defects. However, current planning methods only consider VAs as decision choices. Because VAs and CAs have different activity spaces, planning a joint verification-correction strategy (JVCS) is still challenging, especially for large-size systems. Here we introduce a UCB-based tree search approach to search for near-optimal JVCSs. First, verification planning is simplified as repeatable bandit problems and an upper confidence bound rule for repeatable bandits (UCBRB) is presented with the optimal regret bound. Next, a tree search algorithm is proposed to search for feasible JVCSs. A tree-based ensemble learning model is also used to extend the tree search algorithm to handle local optimality issues. The proposed approach is evaluated on the notional case of a communication system.
\end{abstract}


\begin{center}
  \large\bfseries  
  This work has been submitted to the IEEE for possible publication. Copyright may be transferred without notice, after which this version may no longer be accessible.
\end{center}

\newpage

\section{Introduction}
\label{sec:introduction}
System verification is defined as the process that evaluates whether a system or its components fulfill their requirements~\citep{Engel2010}. System verification is planned and implemented as a verification strategy (VS) that specifies how to implement activities at different developmental phases and on different system configurations~\citep{Salado2018math}. A VS consists of verification activities (VA), each of which is used to identify errors and defects, and correction activities (CA) that correct the identified errors and defects. While VAs includes inspection, analysis, analogy, demonstration, test, and sampling, CAs usually take the form of rework, repair, or redesign, depending on whether the activity encompasses substitution of a faulty part, modification of the product, or modification of the product's design~\citep{Engel2010,international2015quality}. A VS is defined aiming at three objectives: “maximizing confidence on verification coverage, which facilitates convincing a customer that contractual obligations have been met; minimizing risk of undetected problems, which is important for a manufacturer’s reputation and to ensure customer satisfaction once the system is operational; and minimizing invested effort, which is related to manufacturer’s profit”~\citep{salado2015defining}.

Several strategy planning methods have been proposed to design VSs, including decomposition approach~\citep{barad2006optimizing}, set-based design~\citep{xu2019concept}, parallel tempering method~\citep{xu2021parallel}, and reinforcement learning method~\citep{xu2020reinforcement}. All these methods treat only VAs as dedicated decisions that are planned for a system configuration at a development phase (i.e., system state), and CAs are simplified as default actions or even ignored reactively~\citep{xu2021Math}. This simplification may undermine the value of resulting VSs because of the potential suboptimality of CAs. Thus, an extended paradigm of verification planning is presented in our previous paper to include both VAs and CAs as the result of independent decisions. All dependency relationships of VAs and CAs are summarized in that extended paradigm. An order-based dynamic programming (ODP) method is also proposed to find the optimal joint verification-correction strategies (JVCS) that specifies VAs and CAs along a verification process~\citep{xu2021Math}. However, as there are more possible activities, the number of the resulting possible system states increases exponentially~\citep{xu2021Math}. So, it becomes computationally a very difficult task to update the values of all system states. Thus, the ODP method fails to obtain optimal strategies when the system size is large, and an alternative planning method is lacking.

As it is computationally difficult to find the optimal strategy for large systems, we proposed a UCB-based tree search approach for large scale systems in this paper. The proposed method can generate near-optimal JVCSs that fit well with the engineering requirements within certain computational resources. The contributions of this paper are as follows:

First, we presented a search rule based on the upper confidence bound for repeatable bandits (UCBRB). To simplify the verification planning problem, a repeatable bandit model is presented as an extension of the traditional multi-armed bandit problem. Then the UCBRB rule is proposed with the regret theory. The UCBRB rule can be used to find near-optimal strategies of those decision-making problems whose strategies are repeatable. 

Second, we designed a UCBRB1 tree search method to apply the UCBRB rule to verification planning. This method originates from traditional AO* algorithms~\citep{nilsson1982principles} that require accurate admissible heuristic functions, which is hard to define in verification planning. So, we approximate the heuristic functions with the proposed UCBRB rule. The characteristics of system verification, including confidence information, two types of activities, and a value-based model, are also considered to design this method for the search of JVCSs.

Third, we leveraged a tree-based ensemble learning model to handle local optimality issue of the UCBRB1 tree search method. As verification planning is a sequential decision-making problem with dependency, the distributions of state values are not stationary during tree search processes (i.e., concept drift). These concept drifts usually result in local optimality of JVCSs. We trained random forest regression (RFR) with collected samples of system states during the tree search process and used a lower confidence bound of RFR outputs to predict state values as prior information. These prediction values are used to narrow the gap of concept drifts between different system states and help jump out of the local optimum spaces.

The rest of this paper is organized as followed. Section~\ref{sec:literature} reviews the literature about bandit-based methods, tree search methods for sequential decision-making, and tree-based ensemble learning models. Section~\ref{sec:verificationplanning} introduces the basic underlying models to model verification planning problems to underpin the proposed work. Section~\ref{sec:UCBapproach} presents the proposed UCB-based tree search approach. Section~\ref{sec:experimentdesign} uses a demonstrative case to illustrate the performance of the approach. Our conclusions and future directions are summarized in Section~\ref{sec:conclusion}.

\section{Literature Review}
\label{sec:literature}

\subsection{Bandit-based Methods}
\label{subsec:bandit}
The multi-armed bandit problem is a sequential decision model in which an agent needs to decide which arm of $K$ different slot machines to maximize their reward while improving their information at the same time~\citep{bergemann2006bandit}. This problem provides a paradigm of the tradeoff between exploration (trying out each arm to find the best one) and exploitation (playing the arm believed to give the best payoff). The reward of each arm is a random variable $x_k$ with an unknown distribution and the distributions of all arms are independent from each other. This bandit problem aims at finding a policy that determines which bandit to play based on previous trials. The performance of a policy is commonly measured by the agent’s regret~\citep{bubeck2012regret}, which is the expected loss due to not playing the best bandit.

In a seminal paper, Lai and Robbins~\citep{lai1985asymptotically} made a regret analysis to find an asymptotic lower bound on the growth rate of total regret for a large class of reward distributions. The regret bound is $O(\ln n)$, where $n$ is the overall number of plays. Since then, various online policies have been proposed, among which the UCB1 policy developed by Auer et al.~\citep{auer2002finite} is considered the optimal~\citep{kuleshov2014algorithms,huo2017risk}. The UCB1 policy is to play a machine $k$ that maximizes $\bar{x}_k+\sqrt{\frac{2\ln n}{n_k}}$, where $\bar{x}$ is the average reward obtained from machine $k$, and $n_k$ is the number of times machine $k$ has been played so far. That is, this policy can balance the exploitation of the bandit currently believed to be optimal with the exploration of other bandits that currently appear suboptimal but may turn out to be superior in the long run~\citep{browne2012survey}. This UCB1 policy achieves logarithmic regret uniformly over time (not asymptotically) without any prior knowledge regarding the reward distributions. However, the UCB1 is limited by its assumption that the optimal machine is determined by comparing the expected average rewards of all machines. That is, the UCB1 policy would lose its rationality if the expected average reward is not the measure of a machine’s performance.

To solve the tree search problem, Kocsis and Szepesvári~\citep{kocsis2006bandit,kocsis2006improved} proposed the use of UCB1 as tree search policy, which is called the upper confidence bound for trees (UCT). Its formula is
\begin{equation} \label{eq:1}
UCB = \bar{x}_k+D_1\sqrt{\frac{2\ln n}{n_k}},
\end{equation}
where $D_1$ is constant. While the UCT follows the assumption of expected average rewards to determine upper confidence bounds, some variant policies have been proposed as extensions of the UCT to adapt to their domains where the assumption is invalid. We find two domains that are related to system verification in this paper. First, Schadd et al.~\citep{schadd2008single} extended bandit problems to single-player games with perfect information where the expected maximum value rather than the expected average value is used to choose moves. They proposed a single-player MCTS policy (SP-MCTS) that add a third term to the UCT rule: 
\begin{equation} \label{eq:2}
UCB = \bar{x}+D_2\sqrt{\frac{2\ln n}{n_k}}+\sqrt{\frac{x^2-n_k\cdot \bar{x}^2+D_3}{n_k}},
\end{equation}
where $D_3$ is a constant that artificially inflates the standard deviation for infrequently visited nodes. However, as the averaged values are used as the first item to represent the performance, the phenomenon that a good move is hidden by previous records of bad moves may still occur~\citep{bjornsson2009cadiaplayer}. Second, Galichet et al~\citep{galichet2013exploration} studied a risk-aware bandit problem that measures machine’s performance with the conditional value at risk $\alpha$ (CVaR), which is the average of the lowest quantiles of the reward distribution with level $\alpha$. As the goal is to find the machine with maximal CVaR, machines are selected with the best lower confidence bound on their CVaRs, which is called MARAB policy $\widehat{CVaR_k}-D_4\sqrt{\frac{ln(\left \lceil n\alpha\right \rceil)}{\left \lceil n_k \alpha \right \rceil}}$, where $D_4 > 0$ is a constant that controls the exploration vs exploitation tradeoff. As $\alpha$ decreases, the risk-aware bandit problem boils down to a standard max-min optimization problem. However, the calculation of the lowest quantiles results in the storage issue when the bandit problem is of large size, because all previous trials must be recorded to determine the quantiles.

\subsection{Tree Search Methods for Sequential Decision-making}
\label{subsec:treesearchmethods}
Sequential decision-making is a procedural approach to decision-making that aims at finding a policy that maximizes the expected return for all possible initial system states, where earlier decisions influences the later available choices~\citep{barto1989learning}. We refer by sequential decision-making under uncertainty to those problems where the results of a decision follow a random distribution, such as bandit problems, path planning under uncertainty, and system verification. We distinguish between two types of sequential decision-making problems under uncertainty according to whether all following states of a decision should be further planned. If only one following state should be planned after a decision, finding the policy requires determining the set of optimal actions that are connected as a path. In contrast, if it is necessary to plan all following states, the optimal policy can be represented as an AND/OR tree where all actions of a decision have OR relationship and all results of an action have AND relationship. While the first type is widely studied in the fields related to reinforcement learning~\citep{kaelbling1996reinforcement}, the second type has been explored in a variety of research domains, such as system verification~\citep{Salado2018math}, disease diagnosis~\citep{mitra2005bayesian}, and troubleshooting~\citep{khanafer2008automated}.

The key difference between the methods of these two types lies in the calculation of rewards. That is, the reward of a state depends on all following states in the second type of decision-making problems. Until now, two approaches have been proposed to find the policy of the second type. The first approach is heuristic search methods that originate from AO* algorithms for AND/OR trees. Nilsson~\citep{nilsson1968searching,nilsson1982principles} first described a version of AO* for searching AND/OR trees to find a solution in the form of a tree. Martelli and Montanari~\citep{martelli1978optimizing} generalized this algorithm for searching AND/OR graphs to find a solution in the form of an acyclic graph. The graph-search version is more efficient than the tree-search version when the same state can be reached along different paths because it avoids performing duplicate searches~\citep{hansen2001lao}. Hansen and Zilbersterin~\citep{hansen2001lao} proposed a generalization of AO*, called LAO*, to find solution graphs with loops. As the solution of verification planning can be presented as directed acyclic graphs, we only focus on the AO* algorithm for system verification. All these AO* algorithms use admissible heuristic functions to find exact solutions and have been applied to troubleshooting~\citep{warnquist2009planning} and diagnosis of autonomous system~\citep{chanthery2010like}. However, when the admissible heuristic functions are not available, the AO* algorithm cannot be applied.

The second approach is dynamic programming methods (DP) that solve the problem by breaking it down into simpler sub-problems in a recursive manner. DP methods can be classified into exact DP methods and approximate DP methods according to whether the type of solution policies are exact or approximate. Exact DP methods, such as backward induction method, value iteration, and policy iteration, are used to solve exact planning problems~\citep{sniedovich2010dynamic}. They have been applied to solve various fields, such as verification planning~\citep{kulkarni2020verifying,kulkarni2021evaluation}, circuit design~\citep{shachter2010dynamic}, and fault location~\citep{velimirovic2019fault}. However, when the state space is large, exact DP methods fail in finding exact solutions due to the curse of dimensions~\citep{powell2007approximate}. Instead, approximate DP methods can be used to approximate the decision-making process. These approximate methods can be broken down into four classes according to the types of approximate policies: myopic policies, lookahead policies, policy function approximations, and value function approximations~\citep{powell2007approximate}. While myopic and lookahead policies minimize costs for one or several finite time periods, ignoring the effect of a decision now on the later time periods, they are limited if the long-term effects are important. Policy function approximations are applied when the structure of a policy is fairly obvious~\citep{powell2016perspectives}. Because verification strategies have VAs and CAs in this paper, the structure of a policy suffers the dependency relationship between VAs and CAs. So, the fourth type of value function approximations is mostly related to verification planning. Many different approximation models of value functions have been proposed, such as look-up tables of Q learning~\citep{watkins1992q}, support vector regression~\citep{bethke2008approximate}, and neural network~\citep{bertsekas1995neuro}. However, as far as we know, there are no value function approximation methods for the second type of sequential decision-making under uncertainty.

\subsection{Tree-based Ensemble Learning Models}
\label{subsec:ensemblelearning}
Ensemble learning is an effective technique that has increasingly been adopted to combine multiple learning models to improve overall prediction accuracy~\citep{dietterich2000ensemble}. These ensemble techniques have the advantage to alleviate the small sample size problem by incorporating over multiple learning models to reduce the potential for overfitting the training data~\citep{dietterich2000experimental,yang2010review}. Decision trees are commonly used in ensemble learning model because decision trees are sensitive to small changes on the training set~\citep{dietterich2000ensemble}. So, the scope of ensemble learning models is narrowed down to tree-based models. These ensemble learning models have been applied to many fields, such as bioinformatics~\citep{yang2010review}, defect prediction~\citep{matloob2021software}, and remote sensing~\citep{saini2017ensemble}. However, as far as we know, tree-based ensemble learning models have not been used in verification planning.

Some common types of ensemble learning include bagging, boosting, and stocking, which are realized as some basic models, such as random forest (RF)~\citep{ho1995random}, XGBoost~\citep{chen2016xgboost}, LightGBM~\citep{ke2017lightgbm}, and CatBoost~\citep{dorogush2018catboost}. Many studies have been conducted to test the performance of these models and each of them has its own merits. For example, Bentéjac and Csörgő~\citep{bentejac2021comparative} found that CatBoost obtained the best results in generalization accuracy and AUC while LightGBM had the fastest training speed in the studied datasets. Ibrahim et al.~\citep{hancock2020performance} recommended CatBoost algorithm for better prediction of loan approvals and staff promotion. In another study about highly imbalanced Big Data~\citep{jhaveri2019success}, XGBoost was found to be better than CatBoost because of its shorter training time. In addition, the applications of tree-based ensemble learning models are not limited to these basic models. For example, Jhaveri et al.~\citep{ibrahim2020comparison} proposed a weighted random forest along with AdaBoost to predict the success rate of Kickstarter campaigns. Zhang et al.~\citep{zhang2018data} combined RFs with XGBoost to establish the data-driven wind turbine fault detection framework. Zeinulla et al.~\citep{zeinulla2020effective} proposed a fuzzy random forest model to diagnose heart disease with incomplete and dirty datasets. So, the selection of ensemble learning models depends on the characteristics of the research task and datasets.

It is noticeable that ensemble learning has also been used in nonstationary environment, where the underlying data distribution changes over time (i.e., concept drift). Elwell and Polikar~\citep{elwell2011incremental} proposed an ensemble of classifiers-based approach for incremental learning of concept drift. The proposed algorithm collects consecutive batches of data, trains one new classifier for each batch of data it receives, and combines these classifiers using a dynamically weighted majority voting. Later, Yin et al.~\citep{yin2013dynamic,yin2015de2} introduce a comprehensive hierarchical approach called dynamic ensemble of ensembles. It includes two stages. First, component classifiers and interim ensembles are dynamically trained. Second, the final ensemble is then learned by exponentially weighted averaging with available experts. However, all these methods use the weighted averaging method to fuse the information of all components. Thus, they cannot be directly applied in the verification planning that searches for optimal strategies.

\section{Verification Planning Framework}
\label{sec:verificationplanning}

\subsection{System Verification with Bayesian Networks}
\label{subsec:verificationBN}
\begin{figure}
\centering
\subfigure[An Exemplar Network]{
    \begin{minipage}{0.22\textwidth}
    \includegraphics[height=0.75in]{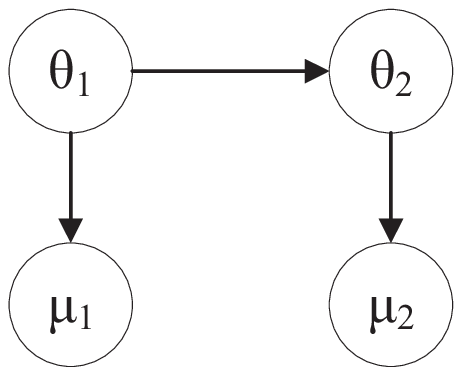}
    \end{minipage}
}
\subfigure[Repair on the BN]{
    \begin{minipage}{0.22\textwidth}
    \includegraphics[height=0.75in]{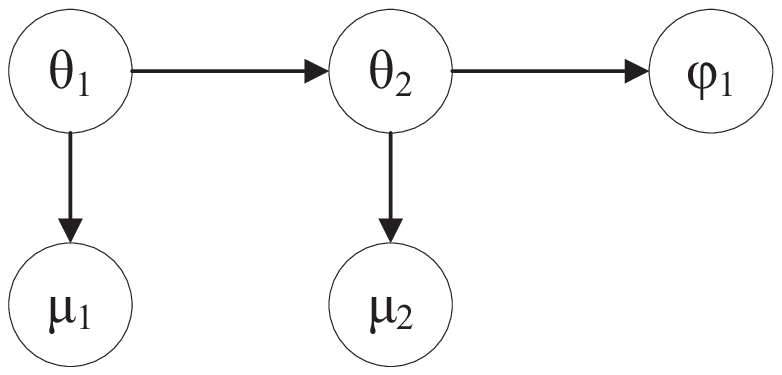}
    \end{minipage}
}
\caption{Illustration of the Bayesian Network Models.} 
\label{fig:BNExample}
\vspace{-2mm}
\end{figure}

We consider that a given system can be decomposed into a set of system elements and assume that the objective of system verification is to verify relevant requirements for these elements. We conceive system verification as a set of tuples of system parameters $\theta_1, \cdots, \theta_I$ about these requirements and the VAs that provide information about such system parameters, denoting the resulting verification evidence of a VA by $\mu_j$ with $j=1,\cdots, J$. Using the modeling framework presented in~\citep{salado2019elemental}, we build a basic system verification model as a BN $\Omega = (\Theta \cup M, E)$, where $\Theta$ is a set of nodes, each of which represents one system parameter, $M$ is a set of nodes that represent VAs, and $E$ is a set of directed edges that connects the different nodes. There are three types of directed edges: $\Theta \Rightarrow \Theta$, $\Theta \Rightarrow M$, and $M \Rightarrow M$, which capture the information dependencies between system parameters, between system parameters and VAs, and between VAs, respectively. In the resulting BN, nodes representing VAs will be treated as observable nodes (those whose node states can be observed directly) and nodes representing system parameters will be treated as hidden nodes (those whose value states cannot be observed directly but are inferred from the values of the observable nodes). For example, consider a computer system that has two parameters, processor speed (denoted by $\theta_1$) and computer speed (denoted by $\theta_2$), and each parameter has its own VA (denoted by $\mu_1$ and $\mu_2$, respectively). The BN can be built accordingly, as shown in Fig.~\ref{fig:BNExample} (a).

Because the interpretation of the information provided by VAs is subjective~\citep{salado2018properties}, we capture the information about system parameters as beliefs. Without loss of generality, all nodes are assumed to be binary (i.e., two node states, such as pass/fail or compliant/non-compliant). The nature of Bayesian analysis, and of BNs by extension, allows for easy removal of this restriction and use of any number of discrete values and even continuous belief distributions~\citep{berger2013statistical}. The specific beliefs of a network node are presented as a conditional probability table (CPT) in this paper. Each CPT summarizes the dependency relationships between a node and all its parent nodes. After all CPTs are elicited as prior distributions of a BN, the impact of a VA on the beliefs is modeled as follows: (1) A verification result $A(\mu_i)$ is collected after executing $\mu_i$ (i.e., an observable node $\mu_i$ is observed); and (2) the posterior distributions of the network nodes are updated by the Bayesian rule.

CAs are defined as those that correct errors or defects that are found during system development~\citep{xu2021modeling}. CAs impact the confidence of system parameters because they affect the system configuration. In our previous study~\citep{xu2021modeling}, uncertain evidence is leveraged to model the effects of CAs on the BN. Three basic types of CAs are modeled with their uncertain evidence: rework, repair, and redesign. For example, when a repair activity is executed to modify a faulty element with parts, processes, or materials that were initially unplanned for that element, it is assumed that repairing the element has impact on the beliefs of the corresponding parameter in the verified system. We can apply virtual evidence to represent the impact of repair on the beliefs. For example, a repair activity is conducted to improve the overall computer speed. As the repair is applied to the system. This piece of evidence is captured as virtual evidence applied on $\theta_2$ directly. The virtual evidence is shown as a virtual node $\varphi_1$ on $\theta_2$ in Fig.~\ref{fig:BNExample} (b). When the uncertain evidence of a CA is collected, the beliefs of a BN are updated in the same way as that of VAs~\citep{xu2021modeling}. That is, when the uncertain evidence of $\varphi_k$ is added to the BN, the confidence of other network nodes can be updated with the Bayesian rule.

\subsection{Verification Planning Problems}
\label{subsec:verificationplanning}

\begin{figure}[htbp]
\vspace{-3mm}
\centerline{\includegraphics[width=7.5cm]{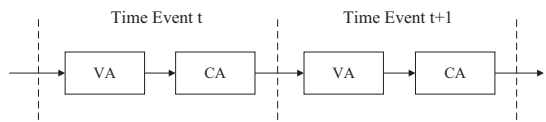}}
\caption{Illustration of Two Time Events of a Verification Process}
\label{fig:timeevent}
\vspace{-3mm}
\end{figure}

\begin{figure}[htbp]
\vspace{-3mm}
\centerline{\includegraphics[width=7cm]{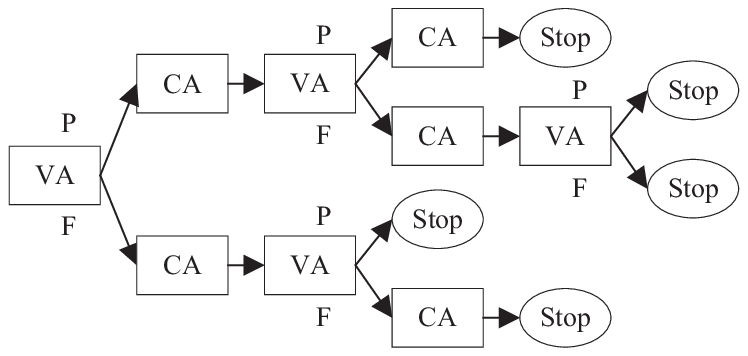}}
\caption{One JVCS example (P: pass; F: fail)}
\label{fig:JVCSExample}
\vspace{-3mm}
\end{figure}

We consider a verification process with time events $t = 1, \cdots, T$. For simplicity, we assume that only one VA and one CA is conducted at each time event and the VA is followed by the CA. This assumption can be relaxed when conducting more than one activity in parallel, which is reserved for future work. In this paper, the paradigm of verification planning is defined as a sequential process of repeating VAs and CAs at $T$ time events, as shown in Fig.~\ref{fig:timeevent}. The solution of verification planning is the assignment of VAs and CAs along a verification process. Because each VA has multiple possible results (e.g., $Pass/Fail$), different combinations of activities exist along the same process and all of these possible combinations can be presented as an activity tree. One example of an activity tree is shown in Fig.~\ref{fig:JVCSExample}. Each path from the root node to a terminal node is called a verification path in this study. In the example in Fig~\ref{fig:JVCSExample}, there are 5 verification paths, and all verification paths share the same initial system state $S_1$. System states are generated along with the collection of activity results at each verification path. At the end of each verification path, the verification process terminates with a certain system state, which is called terminal state (denoted by `Stop’, as shown in Fig.~\ref{fig:JVCSExample}).

At each time event, verification planning consists of assigning a VA and a CA from their own activity spaces, each of which is a set of all eligible activity actions, including the action `NA’ (i.e., No Activity). It is notable that there are two constraints about the activity spaces of VAs and CAs. First, it is unnecessary to repeat any activity again if the previous result of such an activity remains valid. That is, if an activity has been executed and its result can still be used to update beliefs, such an activity is not included in the activity space. Second, implementing a CA can make the existing results of a VA invalid if the result of the VA depends on the corrected parameter. The reason for this is that once a CA changes a system parameter, all existing verification results that depend on such a parameter lose their credibility in deducing accurate posterior beliefs of the system. According to this constraint, it can be inferred that each VA may be executed multiple times if some CAs influence the relevant parameters of such a VA. Therefore, the activity spaces of VAs and CAs depend on the system state at the time the decision is made, and the specific set of activity actions always change along the entire verification process.

To measure the performance of a JVCS, three value factors are considered to calculate the value function. The first value factor is activity costs, which is a fixed amount of financial resources necessary to conduct either a VA or a CA. It is denoted as $C(\mu_j)$ for $\mu_j$ and $C(\varphi_k)$ for $\varphi_k$. For example, if a rework activity is executed to replace a faulty element, corresponding activity costs could include the purchasing cost of new elements and labor fees to replace the elements. The second value factor is failure costs, $C(A(\mu_j)=Fail)$ , which is incurred when the result of a VA is found to be $Fail$. The third value factor is system revenue, $B(\theta_i)$, which is obtained only when a verification process terminates (i.e., reaches a terminal state) and the system is deployed. $B(\theta_i)$ depends on the evolution in confidence that the system is operating correctly as VAs are performed. We consider the system to be deployed only when the confidence level $P(\theta_i=Pass)$ of the target parameter $\theta_i$ at a system state $S_m$ reach or surpass certain thresholds, $H_i$.

For simplicity, all of these value factors are summarized at each terminal state. Each verification path could be stopped in two situations. First, the confidence levels of all target parameters reach their thresholds $\{H_i\}$. Second, the action `NA' is selected when assigning a VA. Consider a JVCS $\Psi$ that starts from a system state $S_1$ and has $w$ verification paths $Z_1, \cdots, Z_W$ and each verification path has a set of system states $\{S_m\}_w$. For a given verification path $Z_w$, the overall value is calculated as: 
\begin{equation} \label{eq:3}
\begin{aligned}
U(Z_w) &= \sum_i B(\theta_i) P(\theta_i=Pass|Z_w)\delta (P(\theta_i=Pass|Z_w) > H_i)\\
&-\sum_j C(\mu_j)-\sum_j C(A(\mu_j)=Fail)-\sum_{k} C(\varphi_k),
\end{aligned}
\end{equation}
where $\delta(\cdot)$ is an indicator function whose value is 1 if the statement is true and 0 otherwise. Because a JVCS $\Psi$ consists of a set of verification paths, the performance of this JVCS $U(\Psi|S_1)$ is calculated as the expected value of a verification process $E(U(Z_w))$, which is the weighted sum of the overall values of all verification paths:
\begin{equation} \label{eq:4}
\begin{aligned}
U(\Psi|S_1) = E(U(Z_w))=\sum_{w} P(Z_w)U(Z_w),
\end{aligned}
\end{equation}
where $P(Z_w)$ is the probability of a verification path $Z_w$ that is calculated by multiplying state transition probabilities (i.e., the probabilities of an activity result) along a verification path:
\begin{equation} \label{eq:5}
\begin{aligned}
P(Z_w) = \prod_{m}{P(A|S_m)},
\end{aligned}
\end{equation}
The verification planning problem is to solve for the optimal JVCS for a given initial system state $S_1$:
\begin{equation} \label{eq:6}
\begin{aligned}
\Psi_{opt} = \arg \max_{\Psi}U(\Psi|S_1),
\end{aligned}
\end{equation}
A summary of the notations used in this paper is shown in Table~\ref{tab:notations}.

\begin{table}[]
\centering
\caption{Summary of Notations}
\label{tab:notations}
\small 
\begin{tabular}[t]{|c|p{11cm}|}
\hline
Symbol                   & Description           \\ \hline \hline
$t$    & Time event, $t=1,\cdots,T$. \\
$\theta_i$ & System parameter, $i=1,\cdots,I$.       \\
$\mu_j$    & Verification activity, $j=1,\cdots,J$. \\
$\varphi_k$    & Correction activity, such as repair, rework, redesign. \\ 
$A$ or $A(\cdot)$  & Activity results. $A(\mu_j)$ is the result of a VA (e.g., $A(\mu_j)=Pass$ or $Fail$). \\
$B(\theta_i)$    & Revenue that depends on system parameter $\theta_i$. \\
$C(\cdot)$    & Cost, including $C(\mu_j)$, $C(\varphi_k)$, and $C(A(\mu_j)=Fail)$. \\
$S_m$                & System state, $m=1,\cdots,M$.          \\
$H_i$                & Threshold value for $B(\theta_i)$.\\
$Z_w$                & Verification path, $w=1,\cdots,W$.\\
$\Psi$            & JVCS, $\Psi=\{Z_w: w=1,\cdots,W\}=\{S_m: 1,\cdots,M\}$.\\ 
$U(\cdot)$                & Value function. $U(S_m)$ is the state value of $S_m$. $U(Z_k)$ is the overall value of $Z_k$. $U(\Psi|S_1)$ is the overall value of $\Psi$ for a given initial system state $S_m$ and $U(\Psi|S_m)=U(S_m)$.\\ 
$X_{k,s}$                & Reward of the $s$th play of the $k$th machine, $k=1,\cdots,K$, which has a distribution $F_k$.\\
$n_{k}$                & Number of plays of the $k$th machine.\\
$n$                & Total number of plays of a bandit problem, i.e., $n=n_1 + n_2 + \cdots + n_k$.\\
$u_k$                & Reward supremum of the $k$th machine.\\
$u_*$                & Maximum reward supremum of all machines.\\
$D_i$                & Constants, including $D_0$,$D_1$,$D_2$,$D_3$,$D_4$,$D_6$,$D_7$,$D_8$,$D_9$,$D_{10}$.\\
$UCB(S_m)$         & Upper confidence bound function of system state $S_m$.\\
$L$         & Lookup table that stores state values and visit counts.\\
\hline
\end{tabular}%
\end{table}

\section{Proposed UCB-based Tree Search Approach}
\label{sec:UCBapproach}

\subsection{Upper Confidence Bound for Repeatable Bandits (UCBRB)}
\label{subsec:UCBRB}
As each verification path has a set of its own system states, all system states of a JVCS can be represented as an AND/OR tree. At any system state whose following system states are not fully explored, verification planning shares the uncertain characteristic with bandit problems. That is, when an activity is selected, the expected reward of this activity follows an unknown distribution. However, verification planning differs from a sequential play of bandit problems. As long as the prior knowledge about the target system is determined, implementing a given JVCS always generates the same expected reward by enumerating all possible verification paths, which means a JVCS is repeatable. So, a repeatable bandit problem (RBP) is investigated in this section first as a simplification of verification planning. 

Consider a K-armed bandit problem where $K$ machines are played sequentially and only one machine is played each time. The reward of a level pull of each machine is represented by a random variable $X_{k,s}$ for $1 \leq  k \leq  K$, where $k$ is the index of a machine. Successive plays of machine $k$ yield rewards $X_{k,1}, X_{k,2}, \cdots$, which are independent and identically distributed (i.i.d.) according to an unknown distribution $F_k$. The support of $F_k$ is $[a_k, b_k]$. The distributions of all machines are independent from each other; i.e., $X_{k,s}$ and $X_{k',s'}$ are independent (and usually not identically distributed) for each $1 \leq k < k' \leq K$ and each $s, s' \geq 1$. Assume that whenever a reward is collected from machine $k$, the player can remember the tricks about reproducing such a reward through controlling the pulling factors, such reaction time point, pull speed, and pull length. So, when the player has collected some pull results of all machines, they can select the machine of the optimal reward and repeat some previous level pull to obtain the same reward. This bandit problem is defined as RBP in this paper. The objective of a RBP is to maximize the sum of rewards earned through a sequence of pulls. 

The main difference between traditional bandit problems and the RBP is the expected reward of each machine. In traditional bandit problems, the reward of a level pull is a random observation of machine $k$. So, the expected reward of machine $k$ is the expectation of $F_k$. However, in RBPs, only the maximum reward of machine $k$ will be considered for further repetition. So, the expected reward of machine $k$ can be represented by the maximum reward after $n_k$ plays of machine $k$, where $n_1 + n_2 + \cdots + n_k  = n$ and $n$ is the total plays of all machines. Other results with lower rewards are not considered for future repetition any more. As each machine of a RBP has its own distribution with a supremum $u_k \leq b_k$, the maximum among the supremum rewards of all bandits is defined as the overall supremum $u_*=\underset{k}{max}(u_k)$. We follow the previous study~\citep{lai1985asymptotically} to define the regret of a RBP as:
\begin{equation} \label{eq:7}
\begin{aligned}
u_*\cdot n-u_k\sum_{k=1}^K E(n_k),
\end{aligned}
\end{equation}
That is, the regret of a RBP is the expected loss due to the fact that the player does not always repeat the optimal play.

For bandit problems, a policy is a strategy that chooses the next machine to play based on the sequence of past plays and obtained rewards~\citep{auer2002finite}. Previous policies, such as the UCB1, work by providing an upper confidence bound for each machine to optimize the accumulated regret. However, as these policies use the expectation of a machine as the target, they are not appropriate choices for RBPs. Thus, we proposed a modified policy called upper confidence bound for repeatable bandits:
\begin{itemize}
\item \textbf{Initialization}: play each machine once;
\item \textbf{Loop}: play machine k that maximizes $x_{k}^{max}+\frac{4ln(n)}{D_0 \cdot n_k}$, where $x_{k}^{max}=max(x_{k,1},x_{k,2},\cdots,x_{k,n_k})$ is the maximum reward obtained from machine $k$, $n_k$ is the number of times machine $k$ has been played so far, $D_0$ is a constant that is determined by the distributions of all machines, and $n$ is the overall number of plays.
\end{itemize}

It is noticeable that $D_0$ is the minimum of ${D_{0,k}}$ and each $D_{0,k}$ is the bound constant of the distribution of each machine, as shown in Lemma~\ref{Lemma}. As the distribution of each machine is unknown, the constant $D_{0,k}$ can be estimated according to presumptive distributions and collected samples. For example, assume the distribution of any machine $F_k$ is a uniform distribution $[a_k, b_k]$. The range $(b_k-a_k)$ can be estimated as $\frac{n+1}{n-1}(x_{k}^{max}-x_{k}^{min})$. So,the estimate of $D_{0}$ is  $\underset{k}{min}\frac{n-1}{(n+1)(x_{k}^{max}-x_{k}^{min})}$.

The regret bound of this policy is summarized as the theorem below. The proof of this theorem is provided in Appendix.
\begin{theorem}
    For all $K > 1$, if the UCBRB rule is run on $K$ machines having arbitrary reward distributions with support in $[0, 1]$, then its expected regret after any number $n$ of plays is at most $O(ln(n))$.
\end{theorem}

\subsection{UCBRB1 Tree Search Method}
\label{subsec:treesearchalgo}
As verification planning can be viewed as a sequential play of RBPs, we use the proposed UCBRB rule to calculate the UCB of the expected reward of a nonterminal system state $S_m$:
\begin{equation} \label{eq:8}
\begin{aligned}
UCB(S_m)=x_{k}^{max}+D_6\frac{ln(n)}{n_k}, D_6 \geq 0
\end{aligned}
\end{equation}
where $x_{k}^{max}$ is the maximum expected reward of $n_k$ JVCSs that starts from $S_m$ (i.e., $\underset{\Psi}{max}(\{U(\Psi|S_m))$\}), $n_k$ is the visit count of $S_m$, and $n$ is the overall visit count of its preceding state $S_{m-1}$. $D_6$ is a constant that depends on the unknown distribution of the state value, which may be obtained through sensitivity analysis in practice. If the range of $x_{k}^{max}$ is larger than 1, the first item may be replaced by $x_{k}^{max}/D_7$, and $D_7$ is a discount constant to normalize the range. With this UCBRB rule, system states are evaluated with the balance between exploration and exploitation. In addition, if the confidence of target parameters reaches the threshold $H_i$ or a `NA' is selected as a VA, the corresponding system state is a terminal state and the reward of this state is deterministic no matter how many more times it is visited. So, the UCBs of these terminal states are simply their maximum reward value and the second item in Eq.~\ref{eq:8} is $0$.

As a verification process consists of multiple time events, the UCBRB rule is insufficient because it only solves the comparison between the activities at a given system state. Therefore, we proposed a UCBRB1 tree search method to make sequential decisions along a verification process. The method consists of a set of loops, with each loop having two stages. First, an AND/OR tree is generated in a forward way where each nonterminal state is expanded as a tip node and an activity is selected to generate its following states. Then, the set of nonterminal states are updated by removing the expanded state and adding nonterminal following states. This expansion step is repeated until the set of nonterminal states is empty. In particular, all feasible activities are determined first according to the activity constraints in Section~\ref{subsec:verificationplanning}. Then each activity is evaluated with the UCB values of all its following states. The activity with the largest value is chosen, as shown in Eq.~\ref{eq:9}:
\begin{equation} \label{eq:9}
\begin{aligned}
\mu_*&=\arg \max_{\mu_j}(-C(\mu_j)-C(A(\mu_j)=Fail)+\sum_{a}P(A(\mu_j)=a)UCB(S_{m+1})|a),\;or\\
\varphi_*&=\arg \max_{\varphi_k}(-C(\varphi_k)+UCB(S_{m+1}|\varphi_k)).
\end{aligned}
\end{equation}
where $S_{m+1}$ means the next system state after conducting the activity and $a$ represents $Pass$ or $Fail$. Second, the node information of all tree nodes, including expected reward and visit counts are updated backwardly from terminal nodes to the root node. That is, the expected value of each node is updated according to expected values of their child nodes in this tree:
\begin{equation} \label{eq:10}
\begin{aligned}
U(S_m)&=-C(\mu_j)-C(A(\mu_j)=Fail)+\sum_{a}P(A(\mu_j)=a)U(S_{m+1}|\mu_*)),\;or\\
U(S_m)&=-C(\varphi_k)+U(S_{m+1}|\varphi_*)).\\
\end{aligned}
\end{equation}
Then the expected rewards are compared with the previous record of this state and saved in the look-up table. If its value is larger than that in the look-up table, the record is updated with the larger one. The visit counts of all tree nodes are added by 1 in the look-up table.

The UCBRB1 tree search method is designed by extending the previous work~\citep{vomlelova2003troubleshooting} from three aspects to improve the efficiency of tree search. First, as all following states of an AND branch must be considered in an AND/OR tree, all nonterminal states of an AND branch are expanded simultaneously in the method. Second, we update the information of all nodes only when the whole AND/OR tree is expanded completely rather than whenever a node is expanded. Third, we use the UCBs rather than an admissible function to represent the heuristic value of a system state.

In addition, two adjustments are made about Eq.~\ref{eq:8} in practice. First, while Eq.~\ref{eq:8} is meaningless if either $n$ or $n_k$ is $0$, the visit counts of all system states are 0 at the beginning. So, we assume that all system states have been visited once before the tree search and, for simplicity, the expected value is set as $None$. Then, Eq.~\ref{eq:8} is equivalent to:
\begin{equation} \label{eq:11}
\begin{aligned}
UCB(S_m)=x_{k}^{max}+D_6\frac{ln(n+1)}{n_k+1},
\end{aligned}
\end{equation}
Second, because all following system states are expanded, the number of tree nodes increases exponentially along with the tree depth, which makes the strategy space very complex. To simplify the strategy space, we add a penalty item in Eq.~\ref{eq:8} to prevent the over-expansion of trees:
\begin{equation} \label{eq:12}
\begin{aligned}
UCB(S_m)=x_{k}^{max}+D_6\frac{ln(n+1)}{n_k+1}-f(m|\Psi),
\end{aligned}
\end{equation}
For simplicity, the penalty item $f(m|\Psi)$ is assumed to be a function of system index $m$ when it is expanded. For example, $f(m|\Psi) = D_8*\left \lfloor m/D_9\right \rfloor$, $D_8=1$, and $D_9=50$. That is, the UCB of a nonterminal state is reduced by 1 every 50 system states. Finally, the UCBRB1 tree search method is summarized in Algorithm~\ref{alg:treesearch}.

\begin{algorithm}
  \caption{UCBRB1 Tree Search Method}
  \label{alg:treesearch} 
  \begin{algorithmic}[1]
    \Inputs{$L=\{ \O\}$: lookup table; $\Psi$: sample tree.}
    \State \parbox[t]{400pt}{Initialize $\Psi=\{ S_1 \}$, where $S_1$ is the initial state.\strut}\label{marker}
    \While{$\Psi$ has some tip nodes}
        \State \parbox[t]{400pt}{Expand all tip nodes of $\Psi$ according to the UCBs of all following states calculated by Eq.~\ref{eq:12}.\strut}
        \State \parbox[t]{400pt}{Add expanded states to $\Psi$.\strut}
        \State \parbox[t]{400pt}{Denote all nonterminal expanded states as tip nodes.\strut}
    \EndWhile
    \State \parbox[t]{400pt}{Update state values and visit counts of all nodes in $\Psi$ according to Eq.~\ref{eq:10}.\strut}
    \State \parbox[t]{400pt}{Update L with the updated state values and visit counts.\strut}
    \If{the expected value $U(\Psi)$ converges}
        \State \parbox[t]{400pt}{Output $\Psi_{opt}$ as the solution.\strut}
    \Else
        \State \parbox[t]{400pt}{Go to~\ref{marker}.\strut}
    \EndIf
  \end{algorithmic}
\end{algorithm}

\subsection{Function Approximation of State Values}
\label{subsec:functionappro}
Even though the proposed UCBRB1 tree search method can solve for a near-optimal JVCS, the tree search process has local optimality issues. That is, the activities of a strategy become fixed at a system state after certain rounds of tree search and it is impossible to explore other possible activities. To handle this issue, we use a function approximation model to approximate state values by generalizing the values of collected system states. Here this model serves two purposes. First, when a new nonterminal state is expanded, the selection of activities suffers from a lack of state information because all following states have not been visited before. However, if a prior state value is added to the calculation of UCBs, some promising action will be selected and the efficiency of tree search will be improved. Second, because the tree search process collect state information incrementally, concept drifts occurs in terms of the distribution of $U(S_m)$. This concept drift can result in the immovability of activity selection, as will be shown in the experiment section. So, the function approximation model is expected to provide some heuristic information to narrow the gap of distributions between different system states. Thus, the tree search process can jump out of local optimum spaces.

\begin{figure}[htbp]
\vspace{-3mm}
\centerline{\includegraphics[width=12cm]{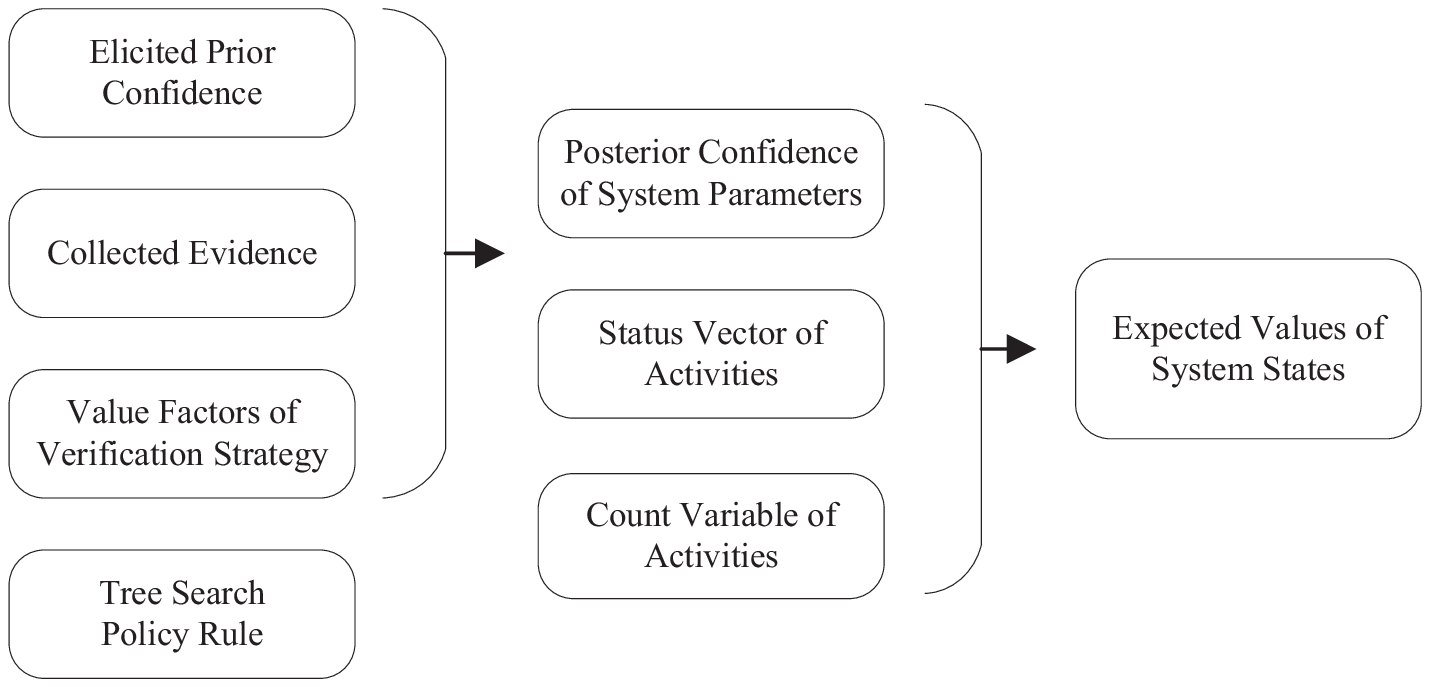}}
\caption{Dependency between Verification Information and State Values}
\label{fig:dependencyplot}
\vspace{-3mm}
\end{figure}

To approximate state values, it is necessary to identify all factors that contribute to the distribution of a state value. During a verification process, there are four types of independent information, including prior confidence of system parameters and activities, collected evidence of activities, value factors of a verification process, and policy rules (e.g., the UCBRB rule). We use three types of variables as the input of the function approximation model to predict state values, as shown in Fig.~\ref{fig:dependencyplot}. First, the posterior confidence values of all system parameters are used because they directly determine whether one system can be deployed. Second, the statuses of all activities, including whether all parameters are corrected and whether valid verification results are collected, are denoted as a list of categorical variables. Third, the counts of all executed VAs and CAs are calculated as two count variables to distinguish node information. These count variables are redundant because they are the sum of all status values.

With the analysis above, we propose to use random forest regression (RFR) to approximate the dependency relationship. The rationale of using RFR for such a function approximation is as follows. First, the input variables are a mixture of continuous, categorical, and count ones. RFR can accept such mixed variables as inputs. Second, all input variables are correlated because they depend on collected evidence. RFR are robust to such redundant inputs. Third, RFR can generate prediction intervals by using the outputs of all decision trees, which can provide a robust approximation of state values. This is important because if an improper activity is selected, all following activities may be vain attempts.

While sample trees are generated continuously along the tree search process, we train RFR models periodically along the tree search process. That is, we divide tree search processes into basic sampling periods and collect tree nodes of sample trees as training datasets at each sampling period. Then a RFR model is built for each sampling period and the output of the latest model are used as the predicted state values. This is explained from two aspects. First, there is a dependency among sample trees because the UCBs are calculated based on all previous state values and visit counts. That is, all training samples are not i.i.d. samples. When the number of sample trees increases, the relationship between training samples becomes increasingly complex. So, models are trained at each period to simplify the dependency between sample trees. Second, because of the concept drift of state value distributions, those built approximation models may lose their generalization accuracy gradually. So, we only use the latest RFR model to predict state values.

We add RFR approximation to extend the UCBRB1 tree search method in two places. First, whenever a sampling period is finished, we train a RFR model based on the lookup table and the sample trees in the sampling period. Because of the randomness of tree search process, the dependency relationship of collected sample trees is random and always change during the tree search process. So, we use RFR models to interpolate the datasets and each decision tree is built with a zero mean squared error. That is, the information of all visited system states are saved accurately in the RFR models. Second, when calculating the UCBs of state values, we use the latest RFR model to predict state values as prior values. In particular, the $k$-th percentile of the outputs of all decision trees is used as the prior state value $\hat{x}_k$. Each prior state value is compared with the state value record in the lookup table. Thus, the UCB formula in Eq.~\ref{eq:12} is extended as:
\begin{equation} \label{eq:13}
\begin{aligned}
UCB(S_m)=max(\hat{x}_k,x_{k}^{max})+D_6\frac{ln(n+1)}{n_k+1}-f(m|\Psi),
\end{aligned}
\end{equation}
where the first item is the maximum between the prior value and the maximum record value.

We also make another two adjustments about data processing to improve the quality of training datasets. First, when each sample tree is generated, all terminal nodes of a sample tree are not included as training samples. The reason is that if `NA' is selected as a VA, this type of terminal nodes suffer larger variance of state values than nonterminal nodes because their following states have not been visited. If parameter confidence reaches the threshold, it is unnecessary to predict state values. Second, besides the tree nodes in the latest sampling period, another set of system states is uniformly sampled from the lookup table with replacement and added to the training datasets. This adjustment is made to reduce the bias caused by the dependency of sample trees. For simplicity, the number of sampled system states is the same as that of sample tree nodes. The extended algorithm is called UCBRB2 tree search method in this paper. 

\section{Experimental Design}
\label{sec:experimentdesign}

\subsection{Problem Description}
\label{subsec:problemdescription}
\begin{figure}[htbp]
\vspace{-3mm}
\centerline{\includegraphics[width=9cm]{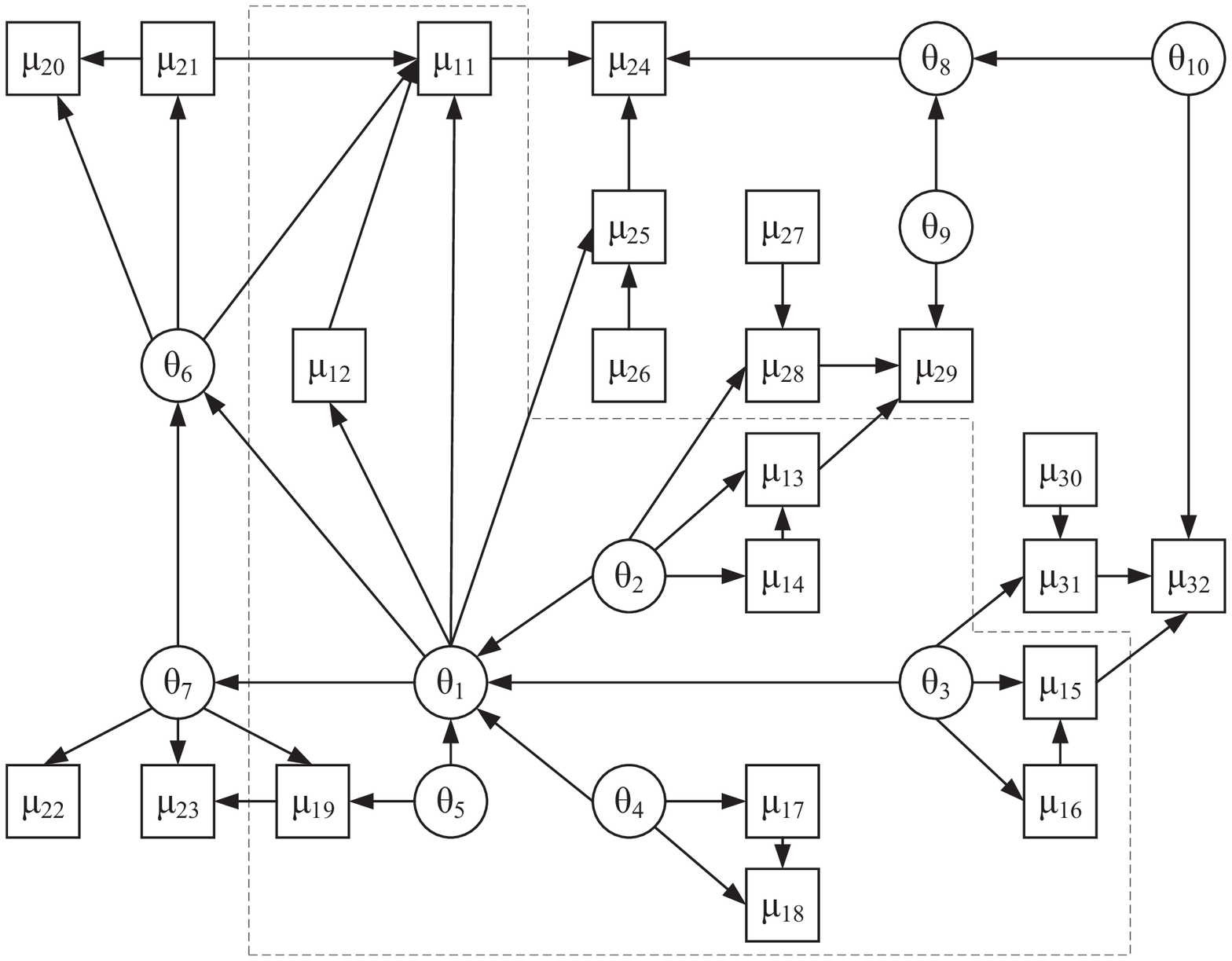}}
\caption{BN of the Optical Instrument (The smaller network outlined by the dash line is used in Scenario 1 and the whole network is used in Scenario 2.)}
\label{fig:BNExperiment}
\vspace{-3mm}
\end{figure}

In this section, we implement the proposed framework to design a JVCS for an optical instrument in a satellite~\citep{salado2019elemental}. The notional instrument has been used to support prior research in verification~\citep{salado2019elemental}. The system parameters of this optical instrument and their possible VAs are modeled as the BN shown in Fig.~\ref{fig:BNExperiment}. System parameters are represented as circle nodes and candidate VAs square nodes. The definitions of the nodes are given in~\citep{salado2019elemental}, hence not presented here. Each node is characterized with its own conditional probability tables (CPT). Their specific values are synthetic and have been generated using the generalized Noisy-OR and Noisy-AND model~\citep{pearl2014probabilistic}, which takes into account the physical meaning of the different modes when estimating their mutual effects for reasonability of the data. While this instrument is verified through a set of these VAs, each parameter $\theta_i$ has CAs to correct potential errors and defects. Without loss of generality, this experiment takes only repair activities $\varphi_k$ as an example of CAs. 

In this experiment, we assume that system revenue is driven by system parameter $\theta_1$. Hence, $\theta_1$ is set as the single target parameter. The threshold for the system deployment rule, $H_1$, is set as 0.90. Cost data have also been synthetically generated in thousand dollar units ($\$1,000$). The revenue $B(\theta_1)$ has been set to $20,000$ so that it provides a balance when making a selection tradeoff between different VAs. In the BN, there is a dependency between $\theta_1$ and all other network nodes. That is, once an activity is executed on any node of these 32 nodes and an activity result is collected, the confidence $P(\theta_1=Pass)$ will change. The activity costs of the different activities, as well as the failure costs of VAs, are provided in Table~\ref{tab:costtable}. Specific values have been generated according to the type of activities defined in~\citep{salado2019elemental}.

\begin{table*}[]
\centering
\caption{Cost Table (Unit: \$1,000)}
\label{tab:costtable}
\resizebox{\textwidth}{!}{
\begin{tabular}{|c|cccccccccccc|}
\hline
Activity & $\mathbf{\varphi_1(\theta_1)}$ & $\mathbf{\varphi_2(\theta_2)}$ & $\mathbf{\varphi_3(\theta_3)}$ & $\mathbf{\varphi_4(\theta_4)}$ & $\mathbf{\varphi_5(\theta_5)}$ & $\varphi_6(\theta_6)$ & $\varphi_7(\theta_7)$ & $\varphi_8(\theta_8)$ & $\varphi_9(\theta_9)$ & $\varphi_{10}(\theta_{10})$ & $\mathbf{\mu_{11}}$ & $\mathbf{\mu_{12}}$ \\ \hline
Activity Cost & 8500 & 5200 & 2000 & 2300 & 1000 & 8500 & 4200 & 6000 & 2300 & 2000 & 3300 & 500 \\
Failure Cost & - & - & - & - & - & - & - & - & - & - & 15000 & 8000 \\ \hline
\hline
Activity & $\mathbf{\mu_{13}}$ & $\mathbf{\mu_{14}}$ & $\mathbf{\mu_{15}}$ & $\mathbf{\mu_{16}}$ & $\mathbf{\mu_{17}}$ & $\mathbf{\mu_{18}}$ & $\mathbf{\mu_{19}}$ & $\mu_{20}$ & $\mu_{21}$ & $\mu_{22}$ & $\mu_{23}$ & $\mu_{24}$ \\ \hline
Activity Cost & 3300 & 3400 & 300 & 300 & 300 & 500 & 100 & 3300 & 3400 & 3400 & 300 & 2400 \\
Failure Cost & 0 & 0 & 0 & 0 & 0 & 0 & 0 & 12000 & 8000 & 2000 & 0 & 10000 \\ \hline
\hline
Activity & $\mu_{25}$ & $\mu_{26}$ & $\mu_{27}$ & $\mu_{28}$ & $\mu_{29}$ & $\mu_{30}$ & $\mu_{31}$ & $\mu_{32}$ & & & & \\ \hline
Activity Cost & 3500 & 2300 & 2400 & 3300 & 2300 & 400 & 300 & 400 & & & & \\
Failure Cost & 12000 & 5000 & 0 & 0 & 0 & 0 & 0 & 0 & & & & \\ \hline
\end{tabular}}
\end{table*}

\subsection{Experimental Method}
\label{subsec:experimentalmethod}
\begin{figure*}[htbp]
\vspace{-3mm}
\centerline{\includegraphics[width=14cm]{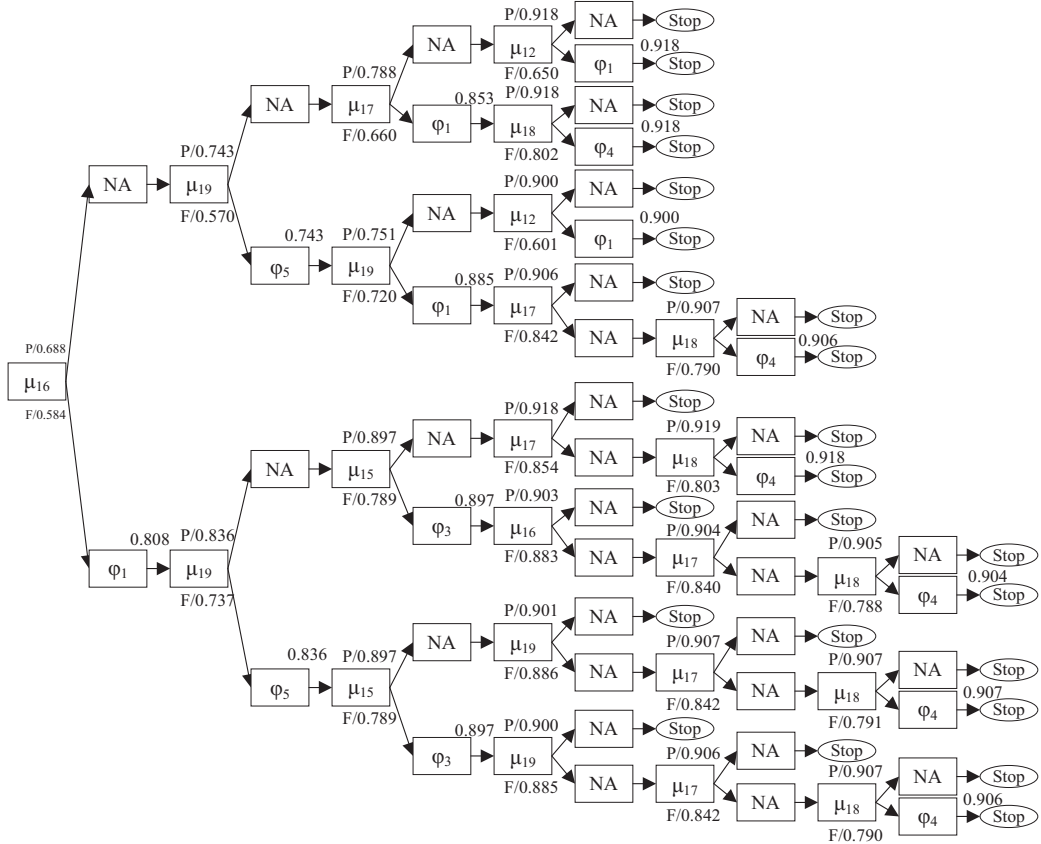}}
\caption{The exact JVCS solution of Scenario 1}
\label{fig:ExactSol}
\vspace{-3mm}
\end{figure*}

With the provided problem and generated data, the whole experiment is realized with Python 3.6 and Bayes Net Toolbox for Matlab~\citep{murphy2001bayes}. Two scenarios are conducted to study the performance of the proposed approach. In Scenario 1, the target network is the smaller one outlined by the dash line in Fig.~\ref{fig:BNExperiment}. With 5 system parameters and 9 VAs in this BN, there are $2\cdot2^5\cdot2^9=1259712$ total system states. The costs items of all activities are shown in bold in Table~\ref{tab:costtable}. To get some intuition about this scenario, we apply the order-based backward induction method first to solve for the exact JVCS. When the backward induction is conducted to calculate the expected values of all states, the optimal activities of all system states are identified to constitute the exact JVCS, as shown in Fig.~\ref{fig:ExactSol}. With 24 verification paths (i.e., the number of terminal states `Stop’), the exact JVCS has 93 tree nodes (excluding terminal states `Stop’). The depth of this JVCS is 13 nodes (i.e., 7 time events). The expected value of this JVCS is $7,788$ while the total running time is $97,925$ sec. 

In Scenario 2, the target network is the whole network in Fig.~\ref{fig:BNExperiment}. With 10 system parameters and 22 VAs in this BN, there are $2\cdot2^{10}\cdot2^{22}=6.43*10^{13}$ total system states. As the number of total system states is too large, the order-based backward induction method is infeasible for this whole network and the exact solution is unknown. However, because the small network is a subset of the whole one, the exact JVCS in Scenario 1 can be used as a reference. So, there is a question whether a better JVCS can be found in the large network, which is discussed later.

In each scenario, the proposed approach is compared with two types of benchmark methods. First, we compare the UCBRB rule with other UCB rules, including the UCT rule (i.e., Eq.~\ref{eq:1}) and the SP-MCTS rule (i.e., Eq.~\ref{eq:2}). As verification planning is such a special problem that we cannot find an accurate admissible heuristic function of all states, the AO* algorithm cannot be applied to solve for AND/OR trees. Algorithm~\ref{alg:treesearch} in Section~\ref{subsec:treesearchalgo} is the only feasible method to find a JVCS as far as we known. So, we combine  Algorithm~\ref{alg:treesearch} with all these UCB rules to compare their effects. For each rule, their constant values are selected through their own sensitivity analyses. We set $D_7=20000$, $D_8=1$, and $D_9=50$. The tree search process is conducted by generating 5000 sample trees. We record the optimal state value every 50 sample trees at the initial state to compare the performance of difference UCB rules.

Second, we compare the proposed methods with a Monte Carlo method. The UCBRB2 tree search method is also tested as an extension of the UCBRB1 one. A RFR model is trained every 3000 tree nodes and another 3000 system states are sampled from the lookup table. Each RFR model consists of 100 decision trees and the 5th percentile of the 100 outputs are used as predicted state values. The hyperparameter `Bootstrap’ is set as $False$ and the hyperparameter `minimum number of samples required to be at a leaf node' is set as 1 to interpolate state samples while other hyperparameters are set as their default values. Finally, a Monte Carlo method is designed to search for a JVCS in a random fashion. As the number of tree nodes can be large if all activities are chosen randomly. A constraint about the total number of tree nodes is added in this Monte Carlo method. That is, the total node number is less than $D_{10} = 50$.

\subsection{Experimental Results}
\label{subsec:expresults}

\subsubsection{Scenario 1}
\label{subsubsec:scenario1}
\begin{figure}[htbp]
\vspace{-3mm}
\centerline{\includegraphics[width=9cm]{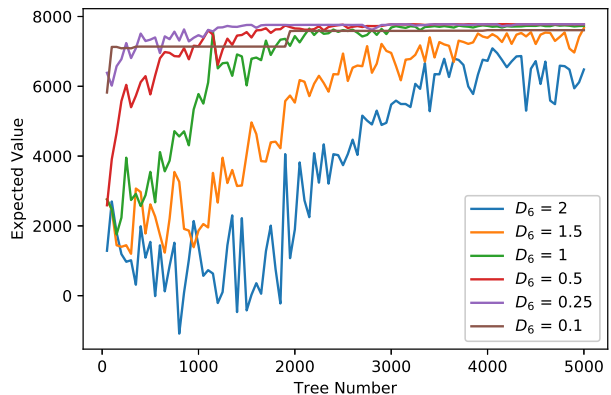}}
\caption{Constant selection for the UCBRB rule}
\label{fig:C0Value}
\vspace{-3mm}
\end{figure}

In the UCBRB rule (given by Eq.~\ref{eq:8}), the constant $D_6$ depends on the unknown distributions of the specific problem. So, it is necessary to determine the constant $D_6$ first for the proposed approach. For simplicity, a set of six possible $D_6$ constants $[0.1, 0.25, 0.5, 1, 1.5, 2]$ is used to find the optimal one. For each constant value, the tree search algorithm in Algorithm~\ref{alg:treesearch} is conducted to test the performance and their trends are shown in Fig.~\ref{fig:C0Value}. It is found that when $D_6$ is 0.5, the expected value reaches the maximum 7780.78 among the 5000 sample trees. The expected values and runtimes of all possible constants are listed in Table~\ref{tab:constantvalues}.

\begin{table}[]
\centering
\caption{Expected value and runtime of all possible $D_6$ constants}
\label{tab:constantvalues}
\resizebox{0.6\textwidth}{!}{
\begin{tabular}{|c|cccccc|}
\hline
$D_6$ & $0.1$ & $0.25$ & $0.5$ & $1.0$ & $1.5$ & $2$ \\ \hline
\hline
Expected Value & 7608.14 & 7777.40 & 7780.78 & 7767.16 & 7670.18 & 7088.53 \\
Runtime & 979.50 & 4510.50 & 7452.68 & 15095.41 & 19899.54 & 22555.04 \\ \hline
\end{tabular}}
\vspace{-5mm}
\end{table}

From Fig.~\ref{fig:C0Value} and Table~\ref{tab:constantvalues}, it can be observed that when $D_6$ decreases, the expected value converges faster. This is because more weights are allocated to the exploitation of existing strategies (i.e., the first item in Eq.~\ref{eq:8}) when $D_6$ decreases. The runtime also decreases become similar trees are exploited. However, if $D_6$ is too small, the tree search process may get stuck in local optimal spaces. For example, when $D_6 = 0.1$, the expected value becomes stable after the 1950th sample tree. Thus, $D_6$ should be large enough given fixed computation resources. In this case, as the limit of tree number is set as 5000, we set $D_6$ as 0.50 for the UCBRB rule in this experiment.

\begin{figure*}
\vspace{-2mm}
\centering
\subfigure[Scenario 1 - UCB rules]{
    \begin{minipage}[b]{0.48\textwidth}
    \includegraphics[width=8.2cm]{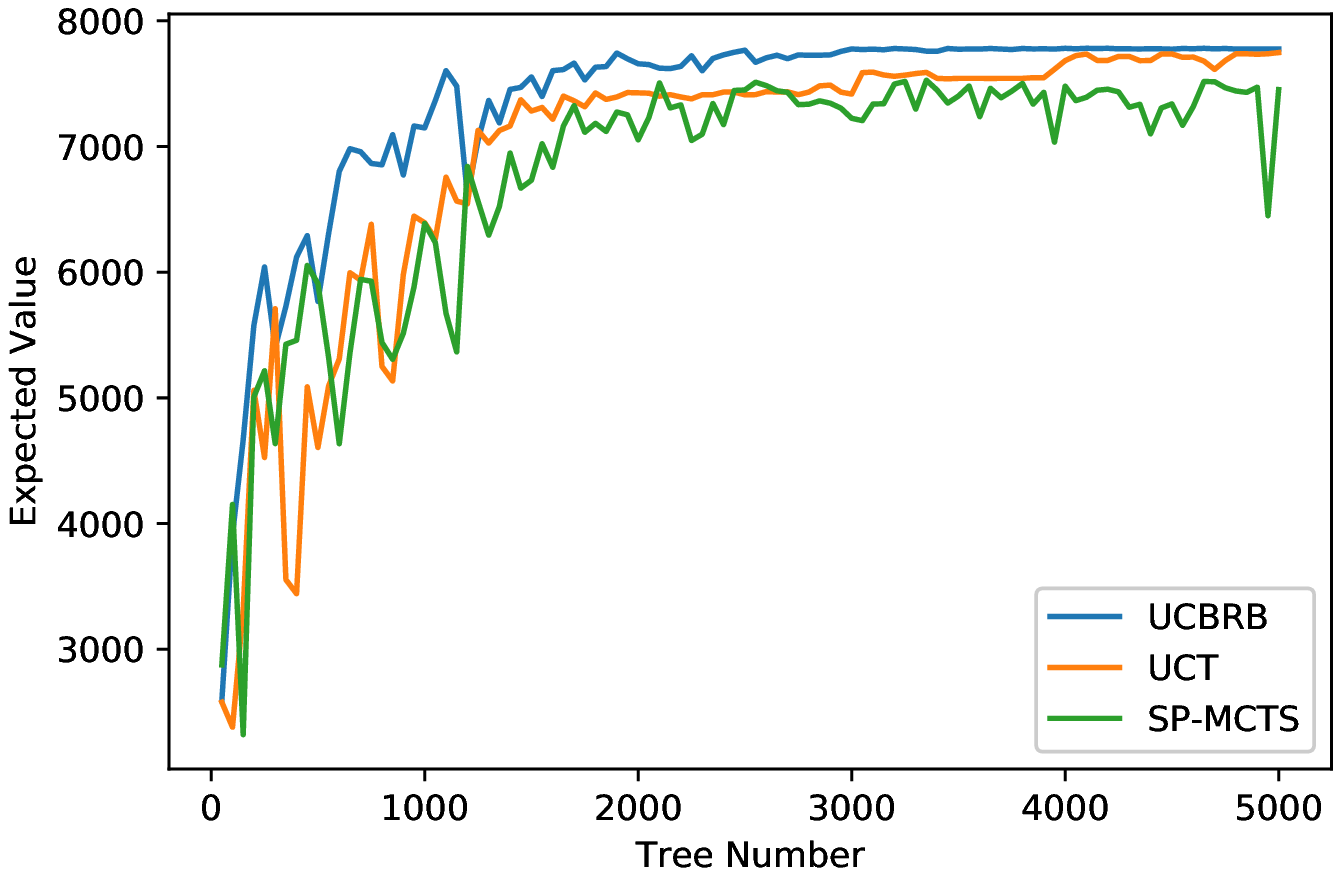}
    \end{minipage}
}
\subfigure[Scenario 1 - Benchmark methods]{
    \begin{minipage}[b]{0.48\textwidth}
    \includegraphics[width=8.2cm]{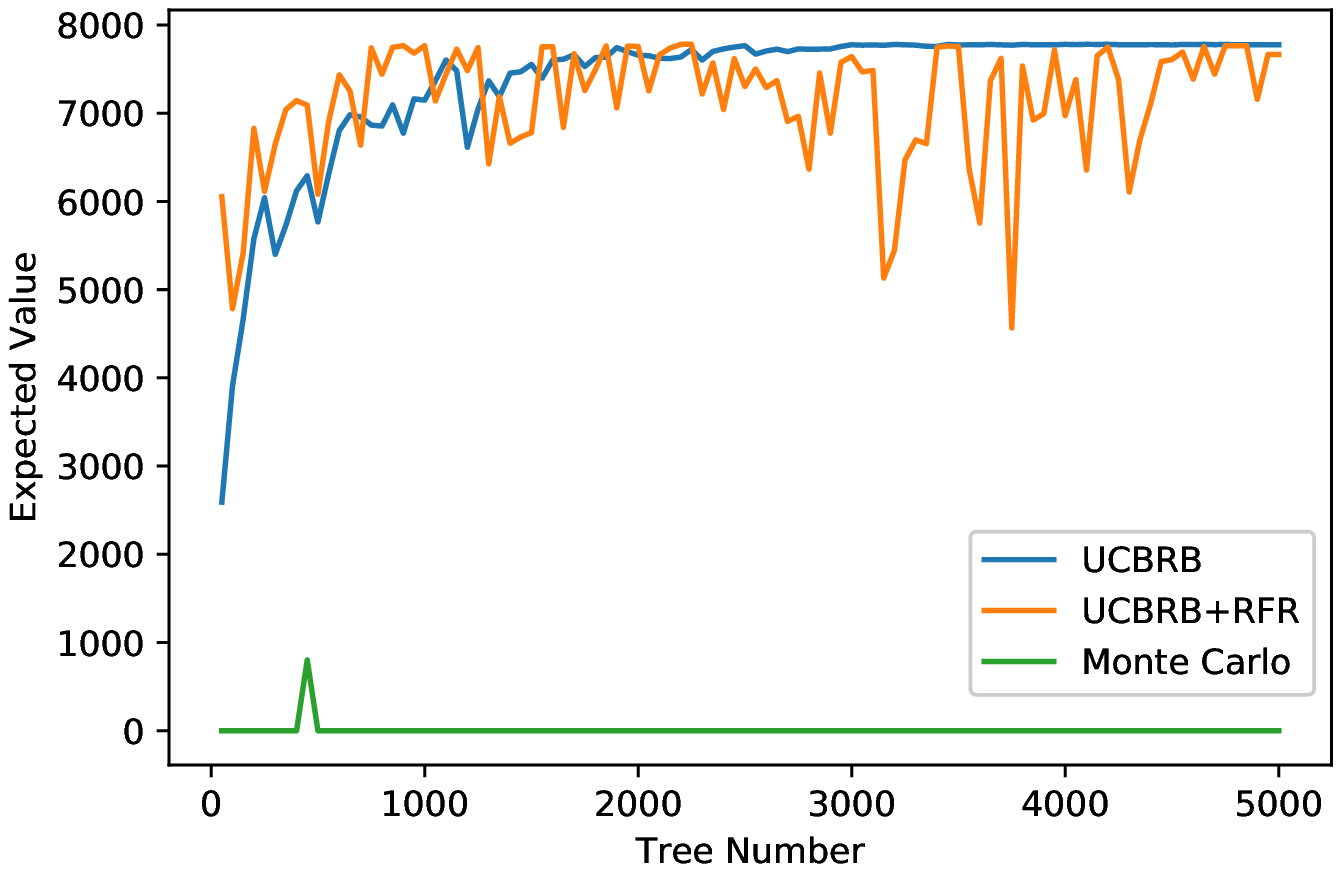}
    \end{minipage}
}
\subfigure[Scenario 2 - UCB rules]{
    \begin{minipage}[b]{0.48\textwidth}
    \includegraphics[width=8.2cm]{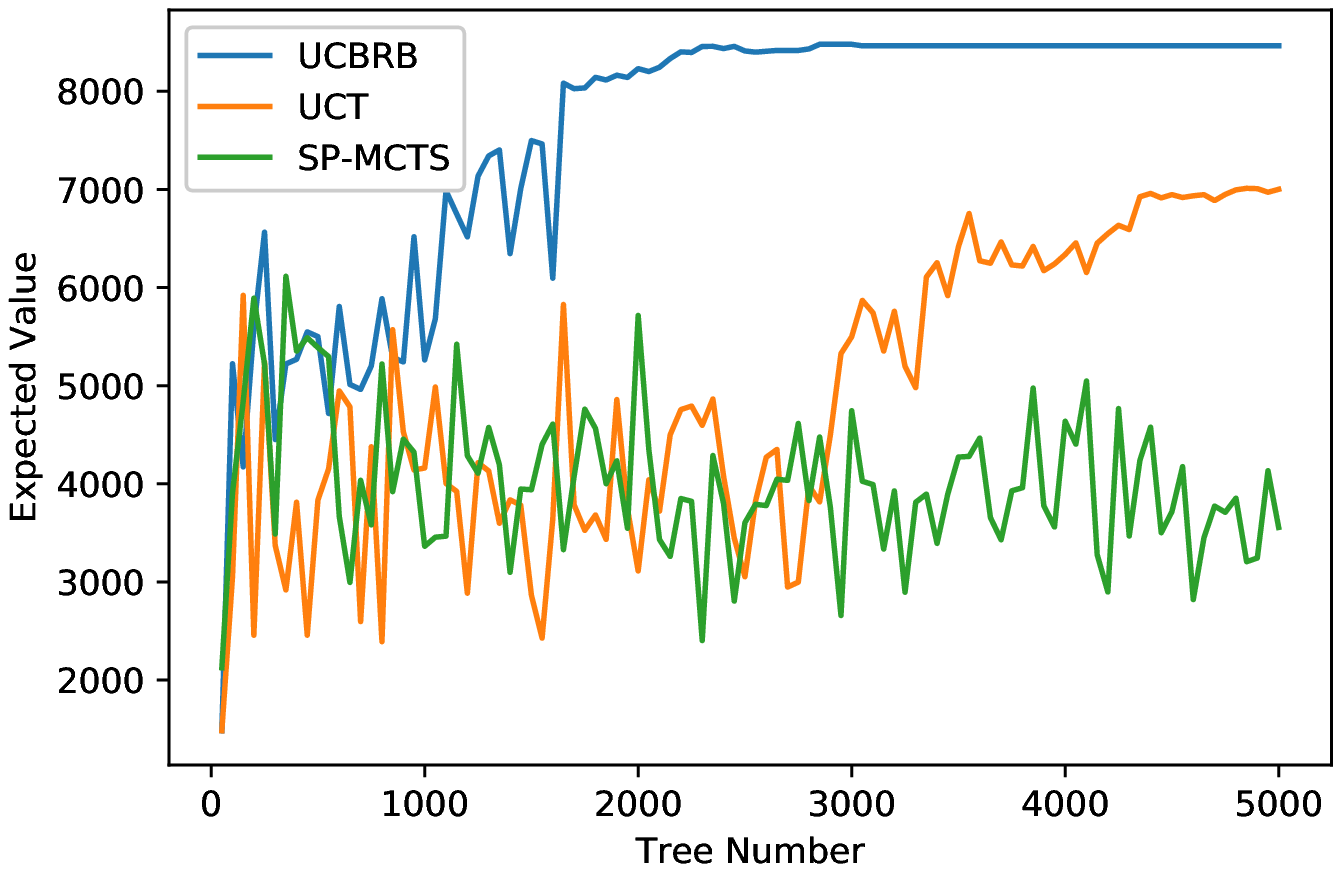}
    \end{minipage}
}
\subfigure[Scenario 2 - Benchmark methods]{
    \begin{minipage}[b]{0.48\textwidth}
    \includegraphics[width=8.2cm]{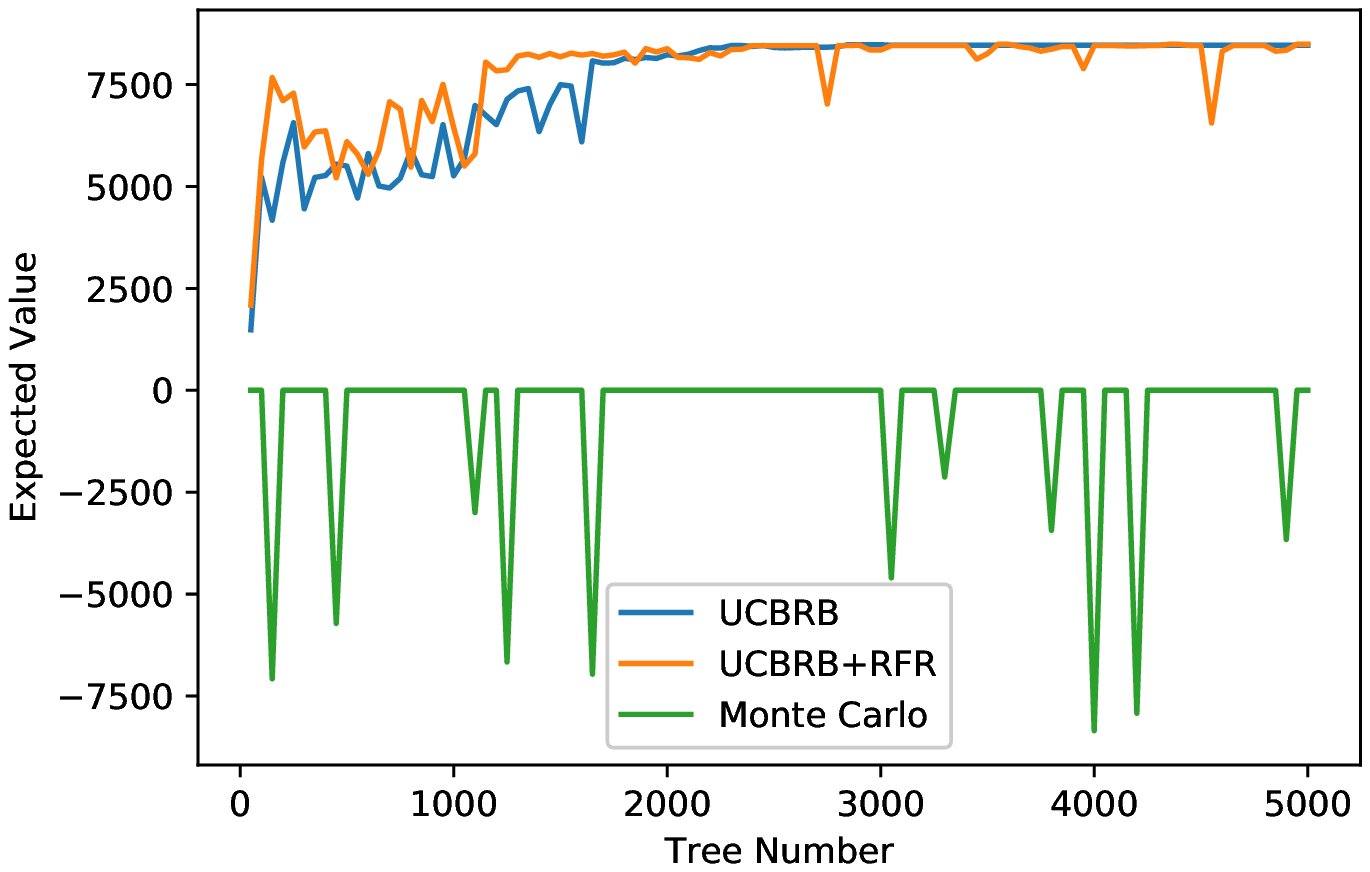}
    \end{minipage}
}
\caption{Plots of All UCB Rules and Benchmark Methods} 
\label{fig:LinePlots}
\vspace{-5mm}
\end{figure*}

Next, three kinds of UCB rules are compared to solve this small network problem. With a similar approach for the $D_6$ value of the UCBRB, the optimal constants of UCT and SP-MCTS is found to be $D_1 = 0.50$, $D_2 = 0.50$, and $D_3 = 10000$. The performance of all UCB rules are shown in Fig.~\ref{fig:LinePlots} (a). It is found that even though both UCT and SP-MCTS rules can find their near-optimal strategies, the UCBRB rule outperforms them slightly after the 4000th sample tree. This is explained by the usage of maximum value as the first item in Eq.~\ref{eq:8}, which make the UCBRB more sensitive to the optimal activities.  It is also found that SPUCB converges faster than UCT. We attribute it to the third item in Eq.~\ref{eq:2} as it adds more weights to those nodes that are visited less frequently. Because there is no significant difference between the optimal expected values of all UCB rules, all UCB rules can be considered for the activity selection in this scenario.

Finally, we also conduct the Monte Carlo method and the UCBRB2 tree search method to search this small strategy space. The Monte Carlo method is conducted to generate 5000 sample trees. Then the UCBRB2 tree search method shares the same $D_6$ value with the UCBRB1 tree search method.
Their performances are shown in Fig.~\ref{fig:LinePlots} (b) and summarized in Table~\ref{tab:performancetable}. It is found that the Monte Carlo method cannot compete with the proposed two methods in terms of the expected values of JVCS, even though it costs the least runtime among all methods. The UCBRB1 tree search method can find the optimal strategy with the highest expected value and its runtime is close to that of the Monte Carlo method.

\begin{figure*}
\vspace{-2mm}
\centering
\subfigure[Scenario 1 - UCBRB1 tree search method]{
    \begin{minipage}[b]{\textwidth} 
    \includegraphics[width=18cm]{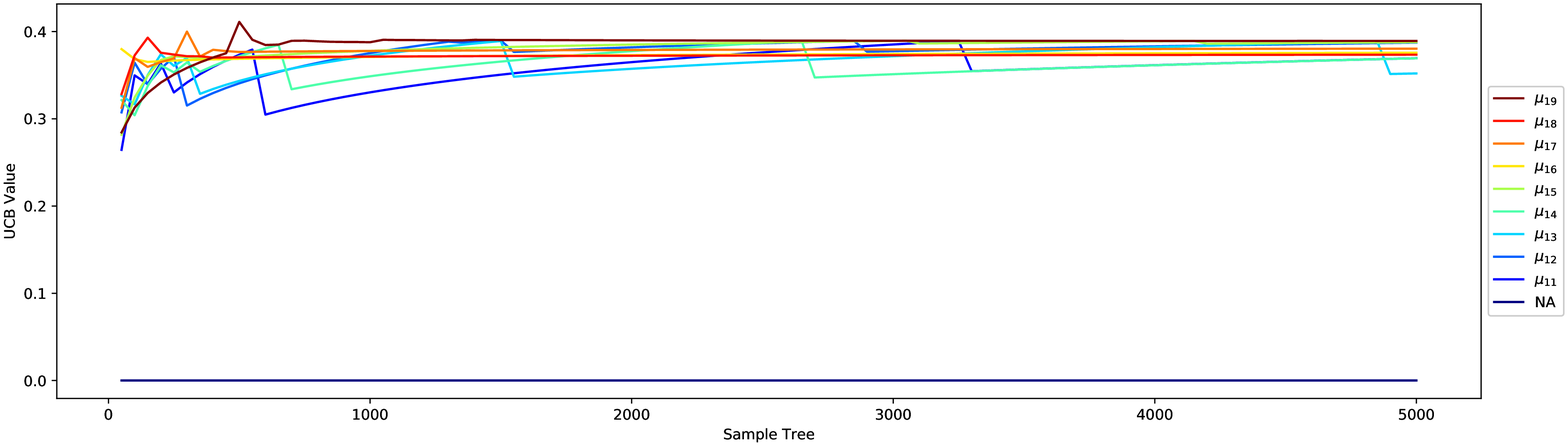}
    \end{minipage}
}
\subfigure[Scenario 1 - UCBRB2 tree search method]{
    \begin{minipage}[b]{\textwidth}
    \includegraphics[width=18cm]{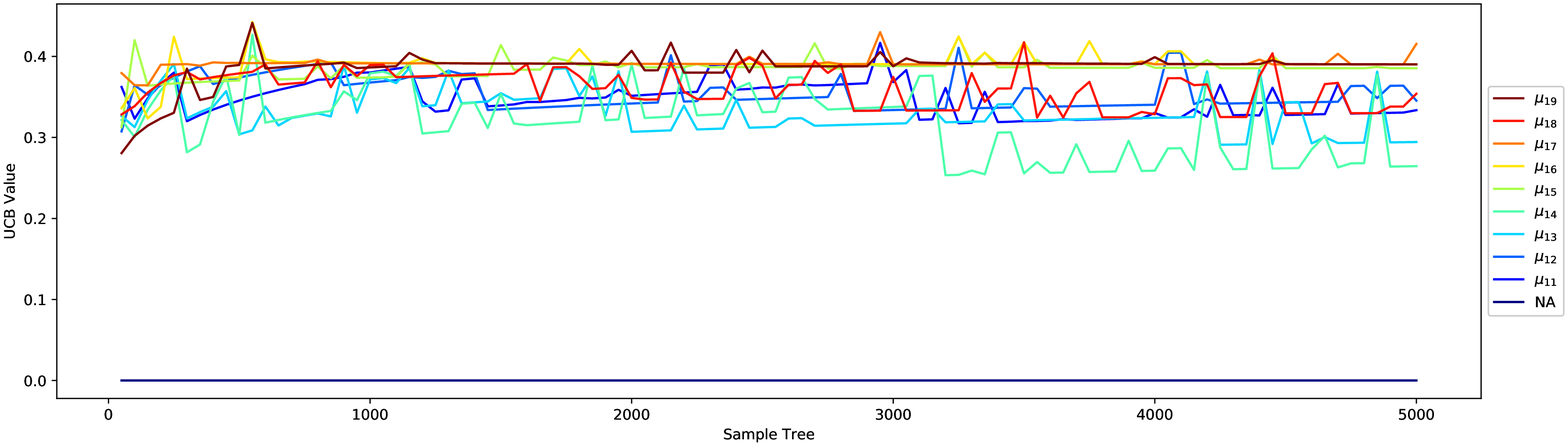}
    \end{minipage}
}
\subfigure[Scenario 2 - UCBRB1 tree search method]{
    \begin{minipage}[b]{\textwidth}
    \includegraphics[width=18cm]{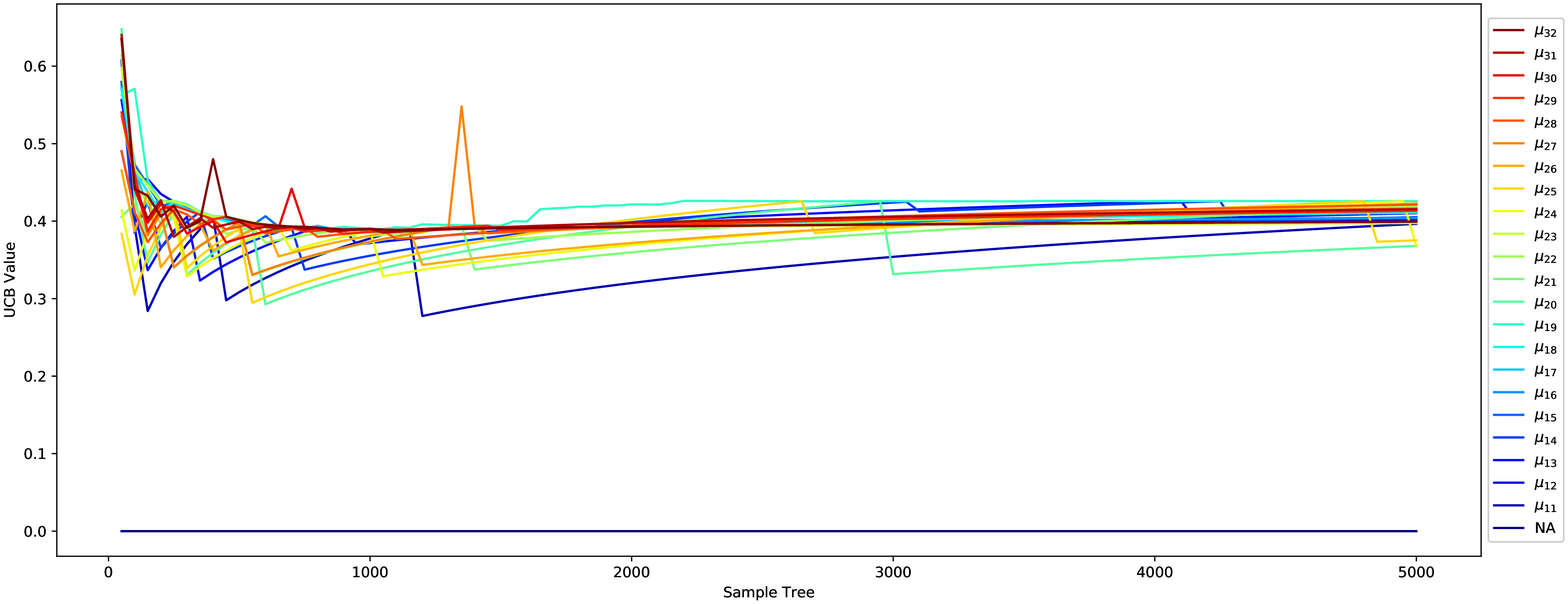}
    \end{minipage}
}
\subfigure[Scenario 2 - UCBRB2 tree search methods]{
    \begin{minipage}[b]{\textwidth}
    \includegraphics[width=18cm]{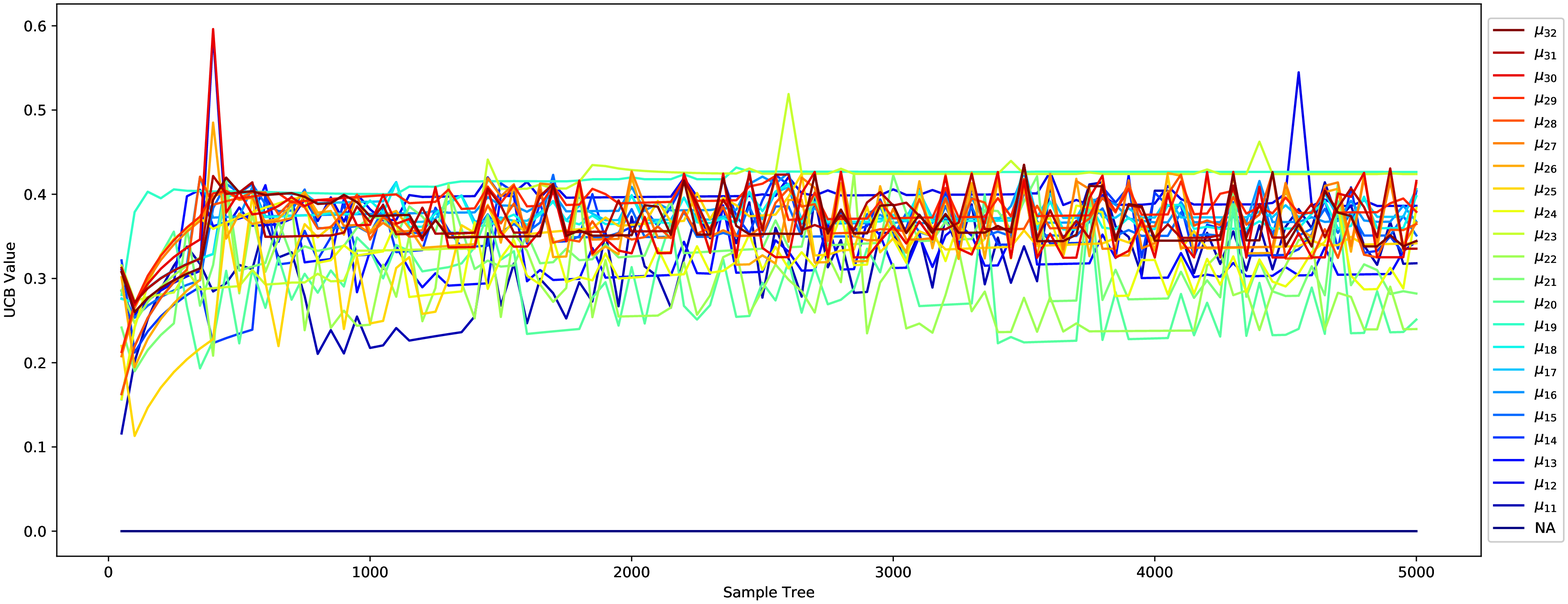}
    \end{minipage}
}
\caption{UCBs of All Activities at the Initial State $S_1$} 
\label{fig:UCBPlots}
\vspace{-5mm}
\end{figure*}

Even though the UCBRB2 tree search method costs the more runtime and its expected value is not the highest, it can solve the local optimality issue that exists in the UCBRB1 tree search method. For simplicity, we use the distribution of all activities at the initial state $S_1$ to study the local optimality issue. If the UCBRB1 tree search method is used, the first activity is always fixed as $\mu_{19}$ after the 500th sample tree, as shown in Fig.~\ref{fig:DistPlots} (a) (i.e., activity immovability). This activity immovability problem is attributed to the concept drift that enlarges the value gap of the first item in Eq.~\ref{eq:8}. So, the UCB of other activities cannot surpass that of $\mu_{19}$, as shown in Fig.~\ref{fig:UCBPlots} (a). The UCBRB2 tree search method solved this local optimality issue by narrowing the value gap with the prior information, as shown in Fig~\ref{fig:UCBPlots} (b). So, more sample trees are allocated to other activities and it is possible to jump out of the local optimum strategy space (i.e., Fig.~\ref{fig:DistPlots} (b)).

\begin{table}[]
\centering
\caption{Performance comparison of all UCB rules and benchmark methods (including UCT, SP-MCTS, Monte Carlo, and ODP)}
\label{tab:performancetable}
\resizebox{0.6\textwidth}{!}{%
\begin{tabular}{|c|cc|cc|}
\hline
               & \multicolumn{2}{c|}{Scenario 1} & \multicolumn{2}{c|}{Scenario 2} \\ \cline{2-5} 
               & Expected Value    & Runtime     & Expected Value    & Runtime     \\ \hline \hline
UCBRB    & 7780.78           & 7452.68     & 8478.43           & 102109.83   \\
UCT           & 7746.82           & 8915.13     & 7011.08           & 281919.13   \\
SP-MCTS          & 7526.81           & 10183.85    & 6114.59           & 227776.50   \\
UCBRB+RFR       & 7780.52           & 12120.16    & 8487.27           & 186804.14   \\
Monte Carlo    & 801.45            & 6797.46     & 0                 & 73234.74    \\
Exact Solution & 7788              & 97925       & -                 & -           \\ \hline
\end{tabular}%
}
\end{table}

\subsubsection{Scenario 2}
\label{subsubsec:scenario2}
In Scenario 2, there are 10 CAs and 22 VAs. As only some extra network nodes are added to the BN, all methods in Scenario 1 can be applied directly. For simplicity, all constants are assigned with the same values as those in Scenario 1. We compare all UCB rules first in this scenario and show their results in Fig.~\ref{fig:LinePlots} (c) and Table~\ref{tab:performancetable}. It is found that the JVCS generated by the UCBRB rule has the maximum expected value 8478.43, while the UCT and SP-MCTS only found their strategies with 7011.08 and 6114.59. In particular, the expected values of the UCT rule do not surpass 6000 until the 3350th sample tree. The reason provided by our analysis is that this large network has more large-cost activities than the small network. So, the mean value is a biased estimate of a state value that makes the estimated UCB less accurate. However, the mean value still has some effect in the long term as it can be found that the expected value gradually reaches 7000. The SP-MCTS rule can find a JVCS with the value 6114.59 at the 300th sample tree. However, once this rule finds the JVCS with the 6114.59, it fails to provide a better JVCS as the value gradually decreases after the 300th sample tree. This is explained by the third item in Eq.~\ref{eq:2} as it is not sensitive to the maximum value when the number of samples is large. Therefore, the UCBRB rule can find a better strategy than other ones in Scenario 2 because it uses the maximum function as a more accurate estimate of state value.

The performances of all three benchmark methods are shown in Fig.~\ref{fig:LinePlots} (d) and also summarized in Table~\ref{tab:performancetable}. It is found that the proposed two methods have found near-optimal JVCSs while the Monte Carlo method fails in providing a JVCS with a positive expected value. The reason is attributed to the large network that has more high-cost activities. So, it is harder to search for a strategy with positive expected values randomly. Instead, estimating the expected value with UCBs can avoid repeating the over-exploration of high-cost activities and yield better strategies with limited tree samples. However, these two proposed methods also cost much more time than the Monte Carlo method. It is attributed to the UCB calculation and the network size. The UCBRB2 tree search method converges faster than the UCBRB1 one in the first 2000 sample trees, even though there is no significant difference between their optimal expected values.

\begin{figure*}
\vspace{-2mm}
\centering
\subfigure[Scenario 1 - UCBRB1 tree search method]{
    \begin{minipage}[b]{\textwidth}
    \includegraphics[width=18cm]{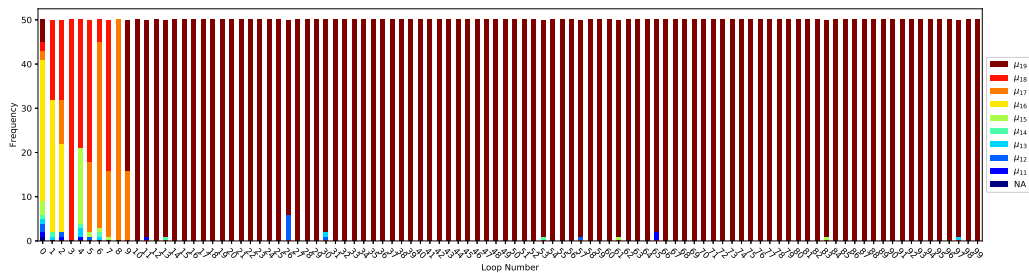}
    \end{minipage}
}
\subfigure[Scenario 1 - UCBRB2 tree search method]{
    \begin{minipage}[b]{\textwidth}
    \includegraphics[width=18cm]{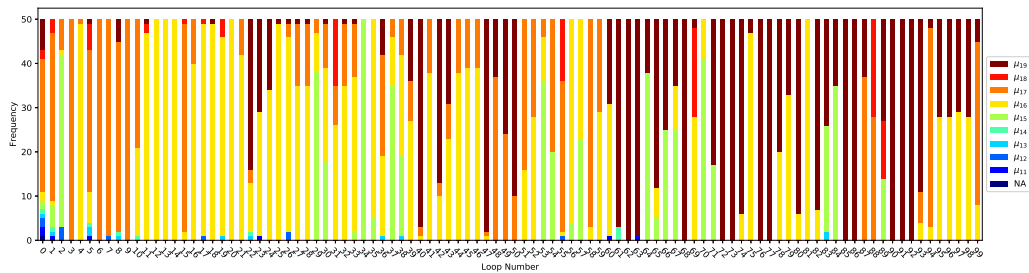}
    \end{minipage}
}
\subfigure[Scenario 2 - UCBRB1 tree search method]{
    \begin{minipage}[b]{\textwidth}
    \includegraphics[width=18cm]{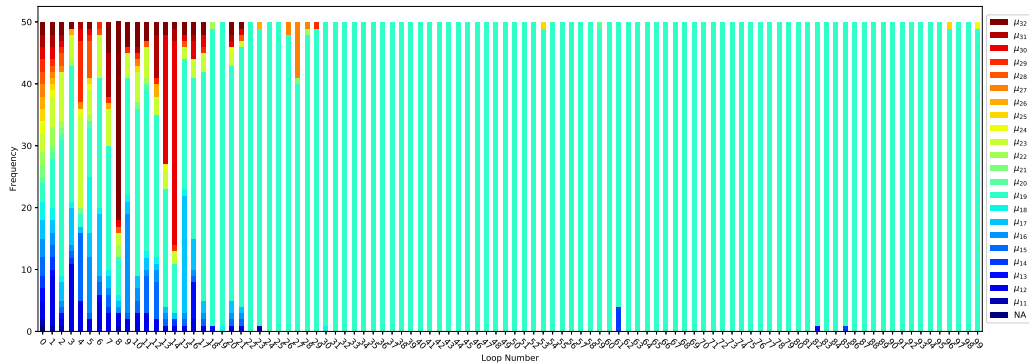}
    \end{minipage}
}
\subfigure[Scenario 2 - UCBRB2 tree search methods]{
    \begin{minipage}[b]{\textwidth}
    \includegraphics[width=18cm]{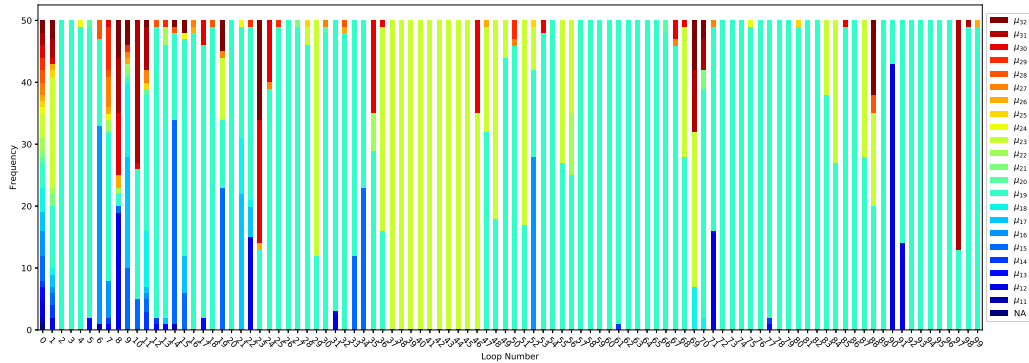}
    \end{minipage}
}
\caption{Distribution of All Activities at the Initial State $S_1$} 
\label{fig:DistPlots}
\vspace{-5mm}
\end{figure*}

The activity immovability problem still occurs when the UCBRB1 tree search method is used. The optimal activity at the initial state is fixed as $\mu_{19}$, as shown in Fig.~\ref{fig:DistPlots} (c). It is caused by the same reason as in Scenario 1 that the value gap of the first item in Eq.~\ref{eq:8} is too large. So, the UCBs of other activities can hardly surpass that of $\mu_{19}$. However, the UCBRB2 tree search method can improve the UCBs significantly so that other activities are allocated with more exploration times. Thus, the tree search process can jump out of local optimum spaces and find a better JVCS. This explains why the JVCS of the UCBRB2 method has the largest expected value 8487.27.

\subsection{Discussion}
\label{subsec:discussion}
The two scenarios in this experiment are designed by expanding the activity set of the same system network. With the proposed UCBRB1 and UCBRB2 tree search methods, a better JVCS is found in Scenario 2, which is attributed to those extra 5 CAs and 13 VAs (i.e., $\varphi_6$ to $\varphi_{10}$, and $\mu_{20}$ to $\mu_{32}$) in Scenario 2. However, this phenomenon does not always occur. We have verified the expected value of the exact JVCS (ref. Fig.~\ref{fig:ExactSol}) in the context of the large network and found its expected value is 76 because the confidence of the target node drops below 0.9 at most terminal states of verification paths in the large network.
Thus, if the costs of all extra activities are very large, it is highly likely that the expected values of near-optimal JVCSs in Scenario 2 are lower than 7,788. Due to some low-cost activities, such as $\varphi_{10}$ and $\mu_{23}$, there are still some opportunities to improve the strategy performance in Scenario 2. Therefore, while finding JVCSs is the task of this research, the effects of all activity costs on JVCSs are not fully explored, which is left as a future work.

During the tree search process, we also found that activity immovability always happens in this verification planning problem, no matter which UCB rule is used.
This problem cannot be solved by adjusting the constant $D_6$, neither.
It is attributed to the concept drift of value distributions. In addition, we assume there are two causes of concept drift. The first one is the lacking of the completeness in tree search processes. If it requires a large number of `Pass' activity results to reach the deployment threshold of the target parameter, the tree of a near-optimal JVCS has a large depth and it cannot be found easily. Instead, only some partial information is collected, which results in concept drift. The second cause is the second item of UCB rules. When the parent visit count is 1 (i.e., $n=0$ in Eq.~\ref{eq:8}), the first item is $None$ and the second item is 0. Then, the optimal activity is `NA' because it has the minimum activity cost $0$. So, the optimal activity of an unvisited state is always `NA', which makes the expected values of its parent tree nodes inaccurate. However, it is noticeable that activity immovability also has a positive effect on the search of JVCS. After the tree search processes find a simple JVCS with a positive expected value, this set of activities is fixed as a core set of activities and its value can be improved by exploitation. Thus, it is unnecessary to eliminate this effect. Instead, it is suggested to improve the UCBs of other promising activities.

While the ensemble learning model is applied to extend the UCBRB1 tree search method, the optimal expected value is not improved significantly. It is mainly attributed to CAs in our view. Because CAs can eliminate errors and defects and reset the statuses of VAs, they can improve the confidence of system parameters fundamentally and reduce the impact of negative activity results. So, if a JVCS can eliminate system errors, the selection of activities at early time events does not matter too much on expected values in this experiment. From the aspect of practice, the UCBRB1 tree search method should be considered first as it can provide a feasible JVCS with less runtime and storage space. However, if it is important to explore other candidate activities, the UCBRB2 method is a better choice because it can handle local optimality issues and solve the exploration-exploitation dilemma much better than the UCBRB1 one.

\section{Conclusion}
\label{sec:conclusion}
This paper has presented a UCB-based tree search approach to solve the verification planning problem. First, we simplify the verification planning problem as a repeatable bandit problem and propose a UCBRB rule. The upper regret bound of this UCBRB rule is also found and proved. Then we propose a UCBRB1 tree search method to apply the UCBRB rule to search for JVCSs. A tree-based ensemble learning model is also used to extend the UCBRB1 tree search method by using RFR models to predict state values. It is found that the proposed UCBRB rule can outperform other UCB rules in the experiment. This advantage is more obvious when the size of the system network increases. The UCBRB2 tree search method can also effectively solve the local optimality issue and handle the exploration-exploitation dilemma better than the UCBRB1 one.

We would like to remark there are three limitations in the proposed methodology. First, the constants of all UCB rules are selected from several possible values in the experiment. The selected value is used as a deterministic value at all system states in the two scenarios. Second, some other parameters and constants of tree search algorithms are not optimized, including discount constant, penalty item, total sample tree number. Third, when training RFR models to predict state values, we set most hyper-parameters with their default values and make some ad hoc adjustments, such as removing terminal states as samples and adding the same number of system states from lookup tables, which are not fully optimized to improve the tree search performance.

In addition, we suggest that this work opens three main future research directions. First, the horizon of verification processes is not fully explored in this study as a penalty item is simply added to restrict the horizon. More mechanisms may be developed to explore the effect of horizon on JVCSs. Second, this work focuses on the tree search given a fixed system network. The selection of possible sub-networks could be explored to simplify tree search processes. The sub-networks may be generated by evaluating the impacts of activities on JVCSs. Third, only RFR models are studied in this work. They are also trained with some intuitive features to approximate system values. Other machine learning methods need to be explored as benchmarks in the future.

\section*{Acknowledgments}
This material is based upon work supported by the National Science Foundation under Grant No. CMMI-1762883.

\section*{Appendix}

\begin{lemma}~\citep[Lemma 4.1]{galichet2013exploration}\label{Lemma}
    Let $v$ be a bounded distribution with support in $[0, 1]$, with $u$ its supremum and let us assume that $v$ is lower bounded in the neighborhood of $a$:
    $$\exists D > 0, \forall \varepsilon >0,P(X\geq a-\varepsilon)>D\varepsilon, with\;X\sim v.$$
    Let $x_1, … x_t$ be t-sample independently drawn from $X$. The maximum value over $x_i, i =1, \cdots, t$ goes exponentially fast to $u$:
    $$P(\underset{1\leq i \leq t}{max} (x_i)\leq u-\varepsilon)>exp(-tD\varepsilon).$$ 
\end{lemma}
    
Proof of Theorem 1:

\begin{proof}
    For each machine $k$, the upper bound of $n_k$ is determined first as $T_k(n)$. Let $h$ be some positive integer, $I_t=k$ is an indicator function, and $X_k^{max}$ be the maximum value over $x_{k,i}, i =1, \cdots, t$. The supremum of $X_{k}^{max}$ is $u_k$ and $u_*=\underset{k}{max}(u_k)$.
    \begin{equation} \label{eq:21}
    \begin{aligned}
    T_k(n)&=1+\sum_{t=K+1}^{n}\{I_t=k\}\\
    &\leq h+\sum_{t=K+1}^{n}\{I_t=k,T_k(t-1)\geq h\}\\
    &\leq h+\sum_{t=K+1}^{n}\{X_*^{max}+\frac{4ln(t-1)}{D \cdot T_*(t-1)} \leq X_{k}^{max}+\frac{4ln(t-1)}{D \cdot T_k(t-1)},T_k(t-1)\geq h\},\\
    &\leq h+\sum_{t=K+1}^{n}\{\underset{0<s<t}{min} X_{*,s}^{max}+\frac{4ln(t-1)}{D \cdot s} \leq\underset{h<s_k<t}{max} X_{k,s_k}^{max}+\frac{4ln(t-1)}{D \cdot s_k}\}\\
    &\leq h+\sum_{t=1}^{\infty}\sum_{s=1}^{t-1}\sum_{s_k=h}^{t-1}\{X_{*,s}^{max}+\frac{4ln(t)}{D \cdot s} \leq X_{k,s_k}^{max}+\frac{4ln(t)}{D \cdot s_k}\}
    \nonumber
    \end{aligned}
    \end{equation}
    
    The inequation $X_{*,s}^{max}+\frac{4ln(t)}{D \cdot s} \leq X_{k,s_k}^{max}+\frac{4ln(t)}{D \cdot s_k}$ implies at least one of the following inequations must hold:
    \begin{equation} \label{eq:22}
    X_{*,s}^{max}\leq u_*-\frac{4ln(t)}{D\cdot s},
    \end{equation}
    \begin{equation} \label{eq:23}
    X_{k,s_k}^{max}\geq u_k+\frac{4ln(t)}{D\cdot s_k},
    \end{equation}
    \begin{equation} \label{eq:24}
    u_* < u_k+2\frac{4ln(t)}{D\cdot s_k},
    \end{equation}
    Because $X_{k,s_k}^{max}$ is always less than $u_k$, Inequation~\ref{eq:23} cannot be true. 
    
    When $s_k \geq \frac{8ln(t)}{D\cdot(u_*-u_k)}$, Inequation~\ref{eq:24} is false because 
    $$u_*-u_k-2\frac{4ln(t)}{D\cdot s_k}\geq u_*-u_k- (u_*-u_k)=0.$$
    
    With Lemma 1, we have
    \begin{equation} \label{eq:25}
    \begin{aligned}
    P(X_{*,s}^{max}\leq u_*-\frac{4ln(t)}{D\cdot s})&\leq exp(-sD\frac{4ln(t)}{Ds})=exp(-4ln(t))=t^{-4} \nonumber
    \end{aligned}
    \end{equation}
    
    So, we have 
    \begin{equation} \label{eq:26}
    \begin{aligned}
    E(T_k(n)) &\leq \left \lceil \frac{8ln(t)}{D(u_*-u_k)}\right \rceil+\sum_{t=1}^{\infty}\sum_{s=1}^{t-1}\sum_{s_k=\left \lceil 8ln(t)/D(u_*-u_k)\right \rceil}^{t-1}P(X_{*,s}^{max}\leq u_*-\frac{4ln(t)}{D\cdot s})\\
    & \leq \left \lceil \frac{8ln(t)}{D(u_*-u_k)}\right \rceil+\sum_{t=1}^{\infty}\sum_{s=1}^{t}\sum_{s_k=1}^{t}t^{-4}\\
    & \leq \frac{8ln(t)}{D(u_*-u_k)}+1+\frac{\pi^2}{6} \nonumber
    \end{aligned}
    \end{equation}
    Thus,
    \begin{equation} \label{eq:27}
    \begin{aligned}
    &u_*\cdot n-u_k \sum_{k=1}^{K}E(n_k)=\sum_{k=1}^{K}(u_*-u_k)E(n_k)\leq \sum_{k=1}^{K}((u_*-u_k)(\frac{8ln(t)}{D(u_*-u_k)}+1+\frac{\pi^2}{6}))=O(ln(t)) \nonumber
    \end{aligned}
    \end{equation}
\end{proof}

\bibliographystyle{unsrtnat}
\bibliography{arXiv}  






\end{document}